\documentclass[%
 reprint,
superscriptaddress,
nofootinbib,
 amsmath,amssymb,
 aps,
floatfix,
]{revtex4-1}

\usepackage{graphicx}% Include figure files
\usepackage{dcolumn}% Align table columns on decimal point
\usepackage{bm}% bold math
\usepackage[colorlinks=true]{hyperref}% add hypertext capabilities
\usepackage[normalem]{ulem}

\begin{document}

\preprint{APS/123-QED}

\title{Astrophysical systematics on testing general relativity with gravitational waves from galactic double white dwarfs}

\author{Shu Yan Lau} \email{sl8ny@virginia.edu}
\affiliation{Department of Physics, Montana State University, Bozeman, Montana 59717, USA}
\affiliation{Department of Physics, University of Virginia, Charlottesville, Virginia 22904, USA}

\author{Kent Yagi} \email{ky5t@virginia.edu}
\affiliation{Department of Physics, University of Virginia, Charlottesville, Virginia 22904, USA}

\author{Phil Arras} \email{pla7y@virginia.edu}
\affiliation{Department of Astronomy, University of Virginia, Charlottesville, Virginia 22904, USA}

\date{\today}

\begin{abstract}
Gravitational waves have been shown to provide new constraints on gravitational theories beyond general relativity (GR), especially in the strong field regime.  Gravitational wave signals from galactic double white dwarfs, expected to be detected by the Laser Interferometer Space Antenna (LISA), also have the potential to place stringent bounds on certain theories that give rise to relatively large deviations from GR in less compact binaries, such as through scalar radiation. Nevertheless, the orbital evolution of close double white dwarf systems is also affected by various astrophysical effects, such as stellar rotation, tidal interactions, and magnetic interactions, which add complexity to the gravity tests. In this work,  we employ the parametrized post-Einsteinian model to capture the leading beyond-GR effect on the signal and estimate the measurement uncertainties using the Fisher information matrix. We then study the systematic error caused by ignoring each astrophysical effect mentioned above on the parameter estimation. Our numerical results show that, to place bounds on the non-GR effects comparable to existing bounds from pulsar observations, tight priors on the mass of the binary and long observation time are required. At this level of sensitivity, we found that systematic errors from the astrophysical effects dominate statistical errors. The most significant effects investigated here are torques from tidal synchronization and magnetic unipolar induction for sufficiently large magnetic fields ($>10^7$G). Meanwhile, even the weaker astrophysical effects from quadrupolar deformations are of a similar order of magnitude as the statistical uncertainty, and hence cannot be ignored in the waveform model. We conclude that the astrophysical effects must be carefully accounted for in the parameter estimation to test gravity with galactic double white dwarfs detected by LISA.
\end{abstract}

\maketitle

\section{\label{Sec:intro}Introduction}

General relativity (GR) has been a successful theory of gravity that satisfies different observational constraints from the solar system to cosmological scales so far. However, there also exists an enormous number of alternative theories of gravity, many of which are well-motivated by fundamental physics or cosmological observations, and are consistent with existing bounds. 
% \sout{In principle, those theories should be treated equally without bias as a possible candidate of the true theory of gravity.} \ky{I felt this sentence is unnecessary.}
Testing GR allows us to narrow down the variety of these alternatives and hence probe the fundamental nature of gravity.

Alternative theories of gravity, or simply non-GR theories, can show deviations from GR within the region of parameter space that are not well-tested. 
One of the most well-studied theories beyond GR is scalar-tensor theories that introduce additional dynamical scalar fields as a source of gravity. Such theories include Brans-Dicke theory \cite{Brans_Dicke_1961}, its generalization that allows neutron stars to spontaneously scalarize \cite{Damour_1992, PhysRevD.87.081506}, $f(R)$ gravity \cite{DeFelice:2010aj} and Horndeski theories \cite{Horndeski:1974wa,Kobayashi:2019hrl}. 
A particular subclass of scalar-tensor theories, known as the screened modified gravity (SMG) \cite{Brax_2012}, features a screening mechanism on the scalar force in dense environments, allowing the theory to satisfy some existing tight constraints \cite{Gubser_2004}.
One can also include dynamical vector fields instead, which introduce a preferred frame and violate Lorentz invariance, with  Einstein-aether theory \cite{ Jacobson_2001,Jacobson:2007veq} and khronometric gravity \cite{Blas:2010hb} as examples. Other theories include Einstein-dilaton-Gauss-Bonnet theory \cite{Kanti_1996} and dynamical Chern-Simons theory \cite{Jackiw_2003, Alexander_2009} that introduce quadratic curvature terms in the action that are coupled to a (pseudo-)scalar field, as well as bigravity theories that contain massive metric fields coupled to the original metric. 
We refer the readers to \cite{Yunes_2013_rev,Yunes:2024lzm,Berti:2015itd, Will_2018} for comprehensive reviews.

Gravitational waves (GWs) allow us to probe gravity in a regime that would be difficult to access with other experiments and observations.
In the weak field regime, precise measurements in solar system tests and binary pulsar observations provide strong constraints on many non-GR theories. Meanwhile, GWs from binary black hole coalescences detected by advanced LIGO and the ground-based detector network provide new probes of gravitation theory in the strong-field regime within dynamical timescales \cite{LIGOScientific:2016lio,Yunes:2016jcc,Berti:2018cxi,Berti:2018vdi,LIGOScientific:2021sio}. One can also look for extra polarizations of the GW signals predicted by some of the non-GR theories \cite{LIGOScientific:2018dkp,Takeda:2021hgo}. By combining electromagnetic signals and gravitational waves, one can measure deviations in the propagation speed of gravitational waves from the speed of light \cite{LIGOScientific:2018dkp}. We refer to e.g.~\cite{Yunes:2024lzm} for a review on the current status of tests of GR with gravitational waves.

Advancements of new observation tools in the future are also anticipated to provide new probes of gravity in different regimes. For the upcoming space-based GW detector, Laser Interferometer Space Antenna (LISA) \cite{2017arXiv170200786A}, much of the literature on tests of GR focuses on using either supermassive black hole binaries or extreme mass ratio inspirals. These sources can improve bounds on various violations of fundamental principles in GR \cite{Perkins:2020tra}, such as a finite mass of the graviton \cite{Berti:2004bd,Yagi:2009zm,Berti:2011jz} and no-hair properties of black holes \cite{Ryan:1997hg,Berti:2005ys,Berti:2016lat}. Multiband observations of stellar-mass black hole binaries with LISA and ground-based detectors will enhance bounds on the scalar dipole radiation \cite{Barausse:2016eii,Perkins:2020tra}, parity violation \cite{Carson:2019rda}, and multiparameter tests of GR \cite{Gupta:2020lxa}. See e.g. \cite{Gair:2012nm} for a comprehensive review on tests of GR with space-based gravitational-wave detectors.

LISA is also expected to detect GW signals from numerous galactic binaries. Binary population synthesis studies show that we can expect up to about 10 thousand double white dwarfs (DWDs) to be resolvable within LISA's frequency band \cite{Lamberts_2019}. These sources span the gravitational wave frequency range between $0.1-100$~mHz, where the lower frequency part 
% \ky{do you agree it's only the low-frequency part of this range that coincides with pulsar binaries?} 
coincides with that from pulsar binaries. They are regarded as quasi-monochromatic signals due to the slow chirping rate, meaning that the phase of the signal carries a limited amount of information about the source. Therefore, these sources are typically not expected to provide a probe of gravity in an entirely new regime. 

Nevertheless, previous studies have shown that DWDs can complement or even improve the current tests on certain non-GR theories in some situations, especially for those effects entering at the negative post-Newtonian (PN) order \cite{Littenberg_2019, Barbieri_2023}. Littenberg and Yunes \cite{Littenberg_2019} first demonstrated the potential of using the galactic binaries detected by LISA to constrain scalar-tensor theories. In particular, they perform a Bayesian analysis and find that when combined with electromagnetic measurements on the masses and radii of the binaries, it is possible to place tighter constraints on the non-GR effects than the current bounds placed by solar system tests and pulsar measurements. Meanwhile, in \cite{Barbieri_2023}, the authors consider the use of both GW measurements of galactic binaries by LISA and by the DECi-hertz Interferometer Gravitational Wave Observatory (DECIGO) \cite{Kawamura_2008} to probe gravity. They focus on testing the time-varying Newton's gravitational constant and show that possible tight constraints can be obtained by DECIGO. However, they also show that sources in LISA's frequency band lead to much weaker bounds than the existing ones without independent constraints on the waveform parameters. It is therefore worth checking in detail how well the GW signals from DWD systems can constrain the non-GR theories.

Testing GR with DWDs relies on the measurements of the chirping of the signal, i.e., the orbital evolution, which depends not only on the GW emission predicted by gravitational theories but also on the detailed properties of the white dwarfs (WD). The latter can include the so-called ``astrophysical factors": the deformation due to spin or tidal field \cite{Poisson_1998}, tidal synchronization torques which transfer angular momentum from the orbit to the WD spins \cite{Benacquista_2011}, and magnetic torques due to systems of currents between the two stars \cite{Lai_2012}, etc. These effects, while Newtonian in origin, can be categorized based on their dependence on orbital separation to enter at different PN orders and can affect the phase in a similar manner as the non-GR effects.  Failure to include these in the waveform model can lead to systematic errors in the measured non-GR effects. 

In this paper, we study the measurability of non-GR effects and the impact of ignoring the astrophysical effects. We employ the parametrized post-Einsteinian (ppE) model \cite{Yunes_2009, Yunes_2010} to study the measurability of non-GR effects in a model-independent way. We consider non-GR effects entering at a variety of PN orders including $-1$PN considered in \cite{Littenberg_2019} for scalar dipole radiation and $-4$PN considered in \cite{Barbieri_2023} for time-varying gravitational constant. By first ignoring the astrophysical effects, we find that the non-GR effects with negative PN orders can be constrained to the same level or even better than the current bounds from binary pulsar observations if there are independent measurements of the mass of the system up to 0.1\% accuracy, an order of magnitude better than current measurements. However, the systematics due to mismodeling of the astrophysical effects, predominantly the corrections from tidal synchronization torques (referred to as the moment-of-inertia factor in Sec.~\ref{Sec:orbit_evolution}), are much larger than the statistical error at this required accuracy. Hence, the astrophysical effects need to be properly modeled in the waveform to place correct bounds on the non-GR effects. We also show that, given reasonable prior constraints on the astrophysical effects, it is possible to include these effects in the model without introducing huge degeneracies between non-GR and astrophysical parameters in the data analysis.

The rest of the paper is organized as follows.
In Sec.~\ref{Sec:orbit_evolution}, we introduce the waveform model and various astrophysical effects in detached close DWDs. In Sec.~\ref{Sec:fisher}, we describe the formalisms used to calculate the measurement errors and the corresponding numerical results. In Sec.~\ref{sec:SMG}, we apply the error analysis with the ppE model on the SMG theory.
In the following, we employ the unit system with $G=c=1$ unless otherwise specified.

\section{\label{Sec:orbit_evolution} Orbital evolution}

The GW frequency evolves over time due to various factors. In detached DWD systems, GR predicts that the leading point-mass contribution to the radiation reaction effect goes as
\begin{align}
    \dot{f}_\text{GR} =& \frac{96}{5\pi \mathcal{M}^2} {x}^{11/2}, \\
    \ddot{f}_\text{GR} =& \frac{33792}{25\pi \mathcal{M}^3} {x}^{19/2}, 
\end{align}
where $\mathcal{M}$ is the chirp mass and ${x} = (\pi \mathcal{M} f_0)^{2/3}$ 
% \ky{Is $\nu$ a standard notation for this quantity? (For example, I think the Living Review article by Blanchet on PN waveforms uses $x$.)} 
is the PN parameter proportional to the square of the relative velocity of a binary in units of the speed of light. The GW frequency $f_0$ is taken to be the initial value at the start of the observation.

For theories beyond GR, the frequency evolves at a different rate. Testing GR involves treating GR as the ``null hypothesis" and looking for deviations from its predictions. The ppE waveform is used to model the deviations due to the leading non-GR effect. In this model, we write the (initial) time derivatives of the GW frequency as 
\begin{align}
    \dot{f}_0 =& \dot{f}_\text{GR}(1+\gamma {x}^{n}), \label{eq:df_approx}\\
    \ddot{f}_0 =& \ddot{f}_\text{GR}\left[1+ \left(1+\frac{2n}{11}\right) \gamma {x}^{n}\right], \label{eq:ddf_approx}
\end{align}
where $\gamma$ is a dimensionless parameter that characterizes the magnitude of the non-GR effect, and $n$ is the post-Newtonian (PN) order relative to the GR effect. The relation between $\gamma$ and the ppE parameter $\beta$ entering in the gravitational wave phase is given by Eq.~(20) of \cite{Tahura:2018zuq} while the expressions for $\beta$ in example non-GR theories can be found in Table I of \cite{Tahura:2018zuq}.

As mentioned above, the astrophysical effects also contribute to the frequency evolution. Up to the leading order of each astrophysical effect, the true waveform would then have the derivatives of the initial frequency
\begin{align}
    \dot{f}_0^\text{(tr)} =& \dot{f}_\text{GR}(1+\gamma {x}^{n}+\Delta), \label{eq:df_tr}\\
    \ddot{f}_0^\text{(tr)} =& \ddot{f}_\text{GR}\left[1+ \left(1+\frac{2n}{11}\right) \gamma {x}^{n} + \left(1+\frac{2 k }{11}\right) \Delta\right], \label{eq:ddf_tr}
\end{align}
where $\Delta$ denotes the astrophysical contributions as described below, and $k$ is the PN order of $\Delta$ with respect to the point-mass contribution, namely $\Delta \propto {x}^k$. We describe below several examples of the astrophysical effects.

\begin{enumerate}
    \item \emph{moment of inertia}:
    
Gravitational wave emission causes the orbital frequency to increase with time, and WD spins will generally start lower than the orbital frequency at large separation (outside the LISA band). As the orbit decays, there will be a time-dependent gravitational force from the companion which will drive fluid motion in the tidally-perturbed star. Any dissipation effect acting on this fluid motion will cause a lag between the tidal forcing and fluid response, and this can give rise to a secular ``tidal synchronization torque" which tries to spin each WD up the orbital frequency. As the tidal torques rob the orbital angular momentum, they increase the rate at which the orbit shrinks.

In the limit that the tidal friction timescale is shorter than the orbital decay timescale, as may happen for close orbits, the WD spins will be nearly synchronized to the orbit and the evolution depends only on a single additional parameter, the combined moment of inertia of the two stars \footnote{When not exactly synchronized, additional parameters for the tidal friction would need to be included.} Following \cite{Benacquista_2011}, this effect is given by
\begin{align}
    \Delta_{I} =& \frac{3 I}{\mathcal{M}^3}{x}^2,
\end{align}
where $I = I_1+I_2$, is the sum of the moment of inertia of the two WDs, (Eq.~(23) in \cite{marsh_2004} provides a fit of the moment of inertia factor as a function of the mass of cold WDs, which we utilize in the following). This contributes to a 2PN correction to the frequency evolution.

\item \emph{spin-induced quadrupole moment}:

Stellar rotation deforms the WD shape, giving rise to a quadrupole moment at the lowest order. The gravity due to this quadrupole moment exerts an additional force on the companion which changes the relationship between orbital separation and frequency. Unlike the synchronization torques, this effect involves a radial force on the orbit and does not rely on dissipation. 

For DWD systems with synchronized spins, the spin-induced quadrupole moment $Q_s$ can be shown to contribute at 5PN order \cite{Benacquista_2011, Poisson_1998}\footnote{The numerical factor here differs from these references as $Q_s$ is assumed to be constant, while it scales as $\Omega^2$ in our derivation under the assumption that the spin remains synchronous all time, where $\Omega$ is the orbital frequency.}, see also Appendix~\ref{app:spinQ}:
\begin{align}
    \Delta_{Q_s} =& \frac{8 \alpha \eta^{2/5}}{\mathcal{M}^4} {x}^{5}, \label{eq:delta_Qs}
\end{align}
where $\eta = m_1m_2/(m_1+m_2)^2$ is the symmetric mass ratio while $\alpha = (Q_{s1}/m_1+Q_{s2}/m_2) / (\pi f_0)^2$. Notice that $\alpha$ is independent of $f_0$ as $Q_{s,i} \propto f_0^2$ to leading order. Through the $I$-$Q_s$ universal relation (originally discovered for neutron stars \cite{Yagi:2013bca,Yagi:2013awa,Yagi:2016bkt}) for WDs \cite{Boshkayev_2017},  it can be written as
\begin{align}
    \bar{Q}_{s,i} = a \bar{I}_{i}^b,
\end{align}
where $\bar{Q}_{s,i} = m_i Q_{s, i}/(\pi I_i f_0)^2$ and $\bar I_i = I_i/m_i^3$ for $i = 1,2$, and $a = 5.255$, $b= 0.4982$. The constants $a$ and $b$ are determined by fitting the numerical values of $I_i$ and $Q_{s,i}$ of a set of cold WD models within the Newtonian framework. The formalism for determining $Q_{s,i}$ follows \cite{Boshkayev_2014, Boshkayev_2017}.

\item \emph{tidally-induced quadrupole moments}:

Another astrophysical contribution comes from the tidal deformation of the WDs. The tidal force from the companion creates a quadrupole moment in the tidally-perturbed star, and the gravity of that quadrupole exerts a force back on the companion. Similar to the spin-induced quadrupole moment, this changes the relationship between orbital frequency and separation, and the effect is independent of dissipation. It causes a 5PN correction as derived in \cite{Flanagan_08,wolz_2021}:
\begin{align}
\Delta_{\Lambda} =& \frac{39 \tilde{\Lambda}}{8 \eta^2}{x}^{5},
\end{align}
where $\tilde{\Lambda}$ is a weighted average of the individual tidal deformability defined as \cite{Wade_2014}
\begin{align}
	\tilde{\Lambda} =& \frac{8}{13}\Big[[(1+7\eta - 31\eta^2) (\Lambda_1 + \Lambda_2) \nonumber\\
	&+\sqrt{1-4\eta} (1+9\eta -11\eta^2)(\Lambda_1 - \Lambda_2) \Big],
\end{align}
and $\Lambda_i = \lambda_i/m_i^5$ is the dimensionless tidal deformability of the WDs. The tidal deformability $\lambda_i$ serves as the proportionality constant of the linear relation between the tidally induced quadrupole moment and the external tidal field from its binary companion. Note that the tidal effect enters at the same PN order as the spin-induced quadrupole moment due to the assumption that both WDs are synchronized. The tidal deformability parameter also enjoys EOS-independent relations with the moment of inertia and the spin-induced quadrupole moment \cite{Boshkayev_2017}.

\item \emph{magnetic field}:

Electromagnetic forces may also change the relationship between orbital separation and frequency, and electromagnetic radiation may alter the rate at which the orbital separation shrinks.

A previous study idealized the electromagnetic interaction of the two magnetized stars as that of two magnetic dipole moments surrounded by vacuum. 
% \sout{The magnetic dipole interaction between magnetized WDs also} 
This interaction induces a change in the orbital decay rate through the magnetic force and the electromagnetic radiation \cite{Ioka_2000, Keresztes_2005,Henry:2023len}. The leading effect enters at 2PN order, assuming the two WDs are bare magnetic dipoles. 
% \sout{ However, in a realistic situation, the plasma around the WDs will form a magnetosphere under the magnetic field, and the interaction between the magnetic dipole moment and the magnetosphere will play an important role.} 
However, in addition to the static currents deep in the highly conductive WDs, plasma in the WD magnetospheres may move to create additional currents and magnetic fields.

The interaction between a moving electrical conductor and a surrounding plasma takes on different forms depending on the parameters of the system. Here we assume that the two WDs are close enough that the ``unipolar inductor" model, first introduced to describe the magnetic interaction between Jupiter and its satellite Io \cite{Goldreich_1969}, is appropriate. Each star raises a wake in the magnetosphere of the other star, which takes on the form of a magnetic flux tube connecting them, with currents running along the tube. These currents close in the resistive atmosphere of the star, giving rise to torques on the star and an equal and opposite torque on the orbit.
This model has been applied to various binary systems including DWDs and neutron star binaries. 

To briefly describe the model, an electromotive force is generated by the orbital motion of the WDs within the magnetosphere and drives a DC current in the system. The orbital energy is dissipated through the resistivity within the circuit. Assuming WD1 is magnetized with a surface magnetic field $B_1$, the model gives \cite{Lai_2012}
\begin{align}
    \Delta_{B} = \frac{5}{64} \eta\zeta_\phi \frac{\Delta \Omega}{\Omega} \frac{\mu_1^2 R_2^2}{\mathcal{M}^6} {x}^{3/2}, \label{eq:DC}
\end{align}
where $\zeta_\phi$ is the azimuthal twist parameter, ranging from 0 to 1, $\Omega$ is the orbital frequency, $\Delta \Omega$ is the difference between the spin frequency and the orbital frequency, and the magnetic moment follows $\mu_1 = B_1 R_1^3$  with $R_A$ representing the radius of the $A$th WD. The above effect enters at 1.5PN order when $\Delta \Omega/\Omega$ is independent of $f_0$. 

In this model, the WDs are assumed to be asynchronous, which is a requirement for the magnetic interaction to impact the rate of orbital decay as seen in Eq.~\eqref{eq:DC}. Meanwhile, we take the systems to be synchronous for the moment of inertia effect as explained earlier. In reality, perfect synchronous is generally not achieved. We make this assumption only to simplify the calculation of the effect from tidal torques, and the corresponding effect on the orbital decay can be considered as the upper limit.
 % \pa{ looks good }

\end{enumerate}

\begin{figure*}[!tp]
    \includegraphics[width=8.6cm]{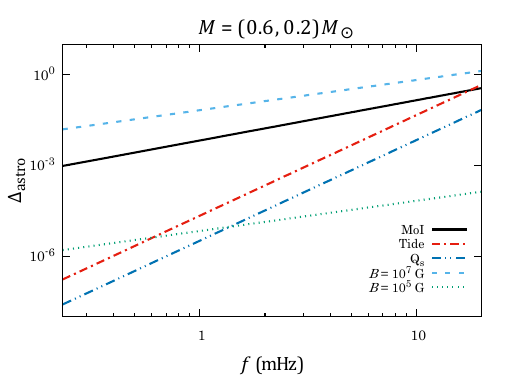}
    \includegraphics[width=8.6cm]{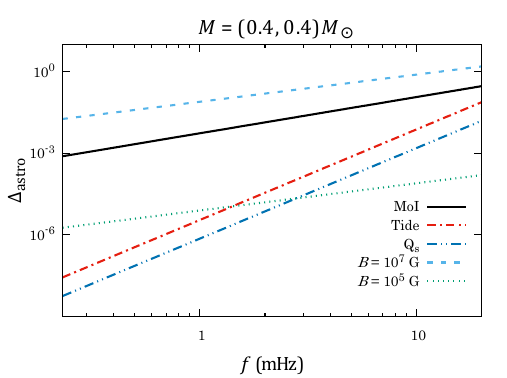}
    \caption{Fractional difference of
astrophysical effects relative to the point-particle contribution on the frequency evolution rate against the gravitational wave frequency of DWD systems of masses (0.6, 0.2)$M_\odot$ (left) and (0.4, 0.4)$M_\odot$ (right) respectively. The effects from moment of inertia (MoI), tidal deformation (Tide), spin-induced quadrupole moment ($Q_s$), and magnetic field ($B = 10^5, 10^7$G) are included.}
    \label{fig:Delta_astro}
\end{figure*}

The above effects enter at different PN orders and therefore scale differently with the frequency of the source. In Fig.~\ref{fig:Delta_astro}, we show the sizes of these effects for two DWD systems of masses (0.6, 0.2)$M_\odot$ and (0.4, 0.4)$M_\odot$. Among $\Delta_I$, $\Delta_\Lambda$, $\Delta_{Q_{s}}$, the moment-of-inertia effect dominates at low frequency due to its 2PN dependence, while the tidal effect becomes more important for very close orbits. The effect from spin-induced quadrupole moment has the same power law dependence as the tidal effect but is about an order of magnitude smaller. The dissipative effect from unipolar induction enters at 1.5PN order and has a $B_1^2$ dependence. For WDs with a strong magnetic field above $10^5$G, it is comparable to or even exceeds that of the rotation effect. In Fig.~\ref{fig:Delta_astro}, we also illustrate the potential of this effect in the strong field case with $B_1 = 10^7$G.  The parameters $\Delta \Omega/\Omega$ and $\zeta_\phi$ are both taken to be 1, denoting maximal asynchronous rate and twist. As discussed in \cite{Lai_2012} and references therein, $\zeta_\phi \gtrsim 1$ would cause the flux tube connecting the circuit to break up. This leads to more complicated situations that are beyond the scope of this paper.

Other possible astrophysical effects include the disturbance due to the external bodies, e.g., the Kozai-Lidov oscillations \cite{Lidov_1962, Kozai_1962} in a hierarchical triple system, or mass transfer between the binary \cite{Verbunt_1988, marsh_2004, Biscoveanu_2023,Yi:2023osk}. The former can affect the orbital elements other than the orbital decay rate, like the eccentricity and inclination, and the latter has a significant impact on the orbital evolution and may cause an out-spiral. Other than that, GR contributions to the orbital motion, like the 1PN effect or the spin-spin interaction \cite{Kidder_1993}, can also lead to deviations from $\dot f_\text{GR}$. These effects have a different origin from the astrophysical effects mentioned above but can still cause a systematic error if not properly included. However, they are not expected to play an important role except for very massive WDs \cite{wolz_2021} and we ignore them for simplicity.

%\section{\label{Sec:fisher} Waveform analysis}
\section{\label{Sec:fisher} Parameter inference}

In this section, we quantify the statistical error caused by the detector noise on the measurement of the non-GR parameter of the ppE model, $\gamma$, as well as the systematic error due to the astrophysical mismodeling of the waveform. We use a Fisher analysis that has been proven to agree well with results from Bayesian Markov-chain Monte-Carlo analyses for tests of GR with DWDs \cite{Littenberg_2019}.

\subsection{\label{ssec:statistical_error} Statistical error}

\begin{figure*}

    \includegraphics[width = 8.6cm]{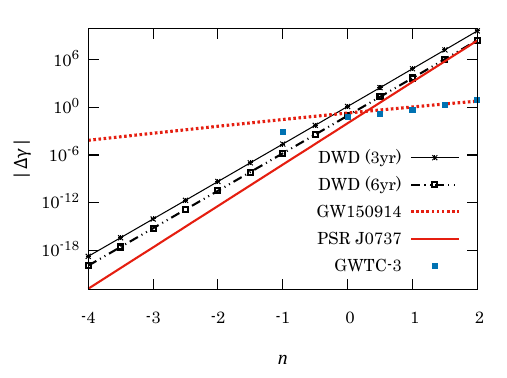}
    \includegraphics[width = 8.6cm]{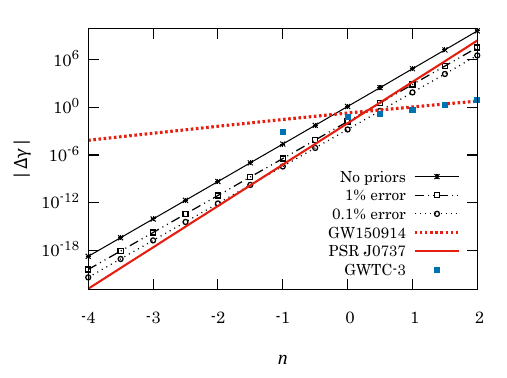}
    \caption{(Left) Statistical error on $\gamma$ for non-GR effects at different PN order $n$ in the ppE waveform model using 3-year and 6-year observations of the DWD listed in the first row of Table \ref{Tab:sources} with LISA. For comparison, we also present constraints set by the binary black hole coalescence event GW150914, the double pulsar binary PSR J0737-3039, and the GWTC-3 data. (Right) Similar to the left panel but with different priors on the mass estimate for a 3-year observation. 
}
    \label{fig:ppE_error}
\end{figure*}

\begin{table*}%[!ht]
\begin{ruledtabular}
\begin{tabular}{ccccccccc}
& SNR (1 yr)& $\phi_0$ & $\psi$ & $\iota$ & $\theta$ & $\phi$ & $f_0$& $\mathcal{M}$\\ 
& & [rad] & [rad] & [rad] & [rad] & [rad] & [mHz]& [$M_\odot$]\\\hline
DWD (sim.) & 95 & $2.2722$ & $0.4413$ & $0.7885$ & $2.1405$ & $4.3525$ & $18.509$& $0.35$ \\
ZTF J1539+5027 & 14.9 & -- & -- & $1.4687$ & $1.9869$ & $-2.7048$ & $4.8217$& $0.30$ \\
NS-WD &  214 & -- & -- & $1.5708$ & $1.7208$ & $4.600$ & $20.0$& $0.53$ \\
XMRI & 19700 & -- & -- & $1.5708$ & $1.7208$ & $4.600$ & $2.0$& $72.5$ \\
\end{tabular}
\end{ruledtabular}
\caption{\label{Tab:sources} Source parameters of the DWD binaries and some other systems considered in Appendix~\ref{app:other_constraints}. The simulated source, DWD (sim.), corresponds to source 4 of Table 1 in \cite{Littenberg_2019}. One of the LISA verification binaries, ZTF J1539+5027 \cite{Burdge_2019}, is also included. For the unspecified angles, we simply take the values as $0$.  
}
\end{table*}

We start by introducing the time-domain waveform model for quasi-monochromatic sources. The strain of the nearly monochromatic GW signal from DWD systems can be described as 
\begin{align}
    h(t) = \mathcal{A}(t) \cos{\Phi(t)}, \label{eq:waveform}
\end{align}
where $\mathcal{A}(t)$ and $\Phi(t)$ are given by
\begin{align}
    \mathcal{A}(t) =& \sqrt{(F_+(t)A_+)^2+(F_\times(t)A_\times)^2},\\
    \Phi(t) = & \Phi_\text{LISA}(t) +\phi_0 + 2\pi f_0 t + \pi \dot{f}_0 t^2 + \pi\ddot{f}_0 t^3/3, \label{eq:phase_taylor}\\
    \Phi_\text{LISA}(t) =& \Phi_D(t) + \Phi_P(t),
\end{align}
where $F_+(t)$, $F_\times(t)$ are the LISA antenna pattern functions (see, e.g., \cite{Cornish_2003}, for the explicit form) while $\phi_0$ and $f_0$ are the phase and frequency at $t=0$. The plus- and cross-polarization amplitudes are $A_+ = A_0 (1+\cos^2 \iota)/2$, $A_\times = A_0 \cos \iota$, where $\iota$ is the inclination angle while $A_0 = 4\mathcal{M}^{5/3} (\pi f_0)^{2/3}/D$ is the dimensionless amplitude that depends on the luminosity distance from the source, $D$, and the chirp mass of the binary, $\mathcal{M}$. In our analysis, $A_0$ is fixed by the signal-to-noise ratio (SNR) of the source. The Doppler modulation and polarization phases are given by
\begin{align}
    \Phi_D(t) =& 2\pi f L \sin\theta \cos{(2\pi f_m t - \phi)}, \\
    \Phi_P(t) =& \tan^{-1} \left(\frac{-F_\times(t)A_\times}{F_+(t)A_+}\right),
\end{align}
where $f_m$ = 1yr$^{-1}$ and $L$ = 1AU. The angles, $\theta$ and $\phi$, are the ecliptic colatitude and longitude respectively. The functions $F_+(t)$ and $F_\times(t)$ depend on $\theta$, $\phi$ and the polarization angle $\psi$.

For these quasi-monochromatic signals, we can approximate the Fisher matrix using a time-domain integration \cite{PhysRevD.57.7089, PhysRevD.69.082005, Shah_2012, Takahashi_2002}:
\begin{align}
    \boldsymbol{\Gamma}_{ij} =& \langle \partial_i h | \partial_j h \rangle \nonumber \\
    =& 4 \text{Re}\left[ \int_0^\infty df  \frac{(\partial_i \tilde{h})^*\partial_j \tilde{h}}{S_n(f)}\right], \nonumber\\
    \approx& \frac{2}{S_n(f_0)} \int_0^{T_\text{obs}} dt \left( \partial_i h\partial_j h\right),
\end{align}
where $\langle .. | .. \rangle$ defines the inner product between two signals, $\tilde{h}$ represents the Fourier transform of the signal $h$, * denotes complex conjugate, $\partial_i$ represents the partial derivative with respect to the waveform parameter $\theta^i$, and $T_\mathrm{obs}$ refers to the observation time. The power spectral density, $S_n(f)$, of LISA follows from \cite{Robson_2019}. The ppE model contains 9 parameters in total: $\boldsymbol{\theta} = \{A_0, \phi_0, \psi, \iota, \theta, \phi, f_0, \mathcal{M}, \gamma  \}$. For signals with long integration time, we can further separate the fast-oscillating part from the slow-evolving part to reduce the computation time \cite{PhysRevD.57.7089}:
\begin{align}
    \boldsymbol{\Gamma}_{ij} \approx& \frac{1}{S_n(f_0)} \int_0^{T_\text{obs}} dt \left( \partial_i\mathcal{A} \partial_j \mathcal{A} + \mathcal{A}^2 \partial_i\Phi \partial_j \Phi\right). \label{eq:fisher_td_approx}
\end{align}
To make this approximation, we have assumed that the fast oscillating parts $\cos{\Phi(t)} \approx \cos{(2\pi f_0 t + \phi_0)}$ and $\sin{\Phi(t)} \approx \sin{(2\pi f_0 t + \phi_0)}$ can be integrated separately from the slowly evolving parts. We further assume that $\int_0^{T_\text{obs}}dt \cos^2{\Phi(t)}= \int_0^{T_\text{obs}}dt \sin^2{\Phi(t)} = T_{\rm obs}/2$ and $\int_0^{T_\text{obs}}dt \cos{\Phi(t)} \sin{\Phi(t)} = 0$.
Similarly, the SNR is given by
\begin{align}
    |h| = \sqrt{\langle h | h \rangle}
    \approx& \sqrt{\frac{1}{S_n(f_0)} \int_0^{T_\text{obs}} dt \mathcal{A}^2}.
\end{align}
Using the ppE waveform, we compute the statistical uncertainties of $\gamma$ by first ignoring the astrophysical effects. For each chosen value of the ppE exponent $n$, we estimate the 1-$\sigma$ uncertainty as the diagonal component of the inverse of the Fisher matrix:
%added ``stat'' on the LHS to match with sys error notation, and 
\begin{align}
    \Delta \theta^i%_\mathrm{stat} 
    = \sqrt{\left(\boldsymbol{\Gamma}^{-1}\right)_{ii}}.
\end{align}

In Fig.~\ref{fig:ppE_error}, we show the statistical uncertainties of $\gamma$ for theories with different $n$ for one simulated detached close DWD systems used in \cite{Littenberg_2019} whose parameters are listed in Table~\ref{Tab:sources} while the fiducial value of $\gamma$ has been set to 0. In the main text, we only consider DWD (sim.) from the table and the others are studied in Appendix~\ref{app:other_constraints}.
The left panel shows results for $T_\text{obs}$ of 3 years and 6 years respectively. The statistical uncertainties have scaling of  $(T_\text{obs})^d$, where $d$ lies between $-3.5$ and $-5$ (see Appendix~\ref{Sec:Error_vs_Tobs}), and hence 6 years of observation provides stronger constraint by an order of magnitude. The constraints on $\gamma$ by some other observations are included for comparison. The first detected binary black hole merger event, GW150914, gives a tighter bound on $\gamma$ for positive $n$ due to the higher frequency of the signal. Meanwhile, the binary pulsar constraint from PSR J0737-3039 provides better bounds than DWDs at almost all PN orders, especially for more negative $n$. The updated bounds on $\gamma$ from the third gravitational wave transient catalog (GWTC-3) \cite{LIGOScientific:2021sio} for $n = -1$ and 0~--~2 are also shown. The mapping between the measured bounds in \cite{LIGOScientific:2021sio} and the ppE parameter $\beta$ is given in Eqs.~(10) and (11) of \cite{Yunes_2016}, while that between $\beta$ and $\gamma$ is described in Sec.~\ref{Sec:orbit_evolution} (below Eq.~\eqref{eq:ddf_approx}).
%\ky{Is it? I see the explanation on the mapping between $\beta$ and $\gamma$, but the bounds from the GW catalog is not exactly on $\beta$...}. 
From the figure, the GWTC-3 provides a stronger bound at $n = -1$ and similar bounds for other values compared to GW150914.

Among the detached DWD systems in the LISA verification binaries \cite{Burdge_2020}, the individual masses of the WDs have been measured up to $\sim$3\% (e.g., J0651+2844 \cite{Hermes_2012}). Assuming we have electromagnetic counterparts or other means of measurement (e.g., GW measurements in other frequency bands \cite{Kinugawa:2019uey}), the GW signal can provide better bounds on $\gamma$. 
In the right panel of Fig.~\ref{fig:ppE_error}, we place priors of different r.m.s. width on the chirp mass (1\%, 0.1\% of the fiducial value respectively) following \cite{Littenberg_2019}. Because we are using a Fisher analysis, we impose Gaussian priors through the replacement $\boldsymbol{\Gamma}_{ij} \rightarrow \boldsymbol{\Gamma}_{ij}+\delta_{ij}/\sigma_i^2$ \cite{Poisson:1995ef,Berti:2004bd}, where $\delta_{ij}$ is the Kronecker delta function and $\sigma_i$ is the r.m.s. width of the $i$th parameter. 
Note that the repeated indices are not summed over. We then computed the statistical error of $\gamma$ for an observation time of 3 years. We find that, as expected, imposing tighter priors can improve the bounds on $\gamma$. 
In particular, a precise measurement of the chirp mass up to 0.1\% would result in more stringent constraints than the pulsar for some $n$. This result is similar to that in \cite{Littenberg_2019} for the constraints on the coupling parameters of a specific class of scalar-tensor theories.

\subsection{\label{ssec:systematic_error} Systematic error}

The astrophysical effects introduced in Sec.~\ref{Sec:orbit_evolution} cause deviations in $\dot f_0^\text{(tr)}$ and $\ddot f_0^\text{(tr)}$ (Eqs.~\eqref{eq:df_tr} and \eqref{eq:ddf_tr}) from the approximate model used in Sec.~\ref{ssec:statistical_error}. In this subsection, we focus on estimating the systematic errors on $\gamma$ caused by neglecting these effects in the waveform model following \cite{Cutler_2007}. 
We first briefly review the formalism and apply it to each astrophysical contribution to the waveform.

We assume that the true waveform signal that includes all contributions from the astrophysical effects, as well as the non-GR effects, is given by 
\begin{align}
    h_0(t) = \mathcal{A}_0(t) \cos{\Phi_0(t)},
\end{align}
while the approximate waveform model we use for parameter estimation is 
\begin{align}
    h_\text{A}(t) = \mathcal{A}_\text{A}(t) \cos{\Phi_\text{A}(t)}.
\end{align}
The true waveform parameters are denoted by $\boldsymbol{\theta}_0$. In particular, $\Phi_0(t)$ is expressed as Eq.~\eqref{eq:phase_taylor} with $\dot{f}_0$ and $\ddot{f}_0$ replaced with $\dot{f}_0^{\text{(tr)}}$ and $\ddot{f}_0^{\text{(tr)}}$ while the approximate waveform is described in Eq.~\eqref{eq:waveform}. Using the latter waveform model, the measured parameters, denoted by $\boldsymbol{\theta}_\text{A}$, would then have a systematic shift from the true values estimated by the formula\footnote{The original formula in \cite{Cutler_2007} is in frequency domain while we extended it to time domain.}
% \begin{align}
%     \Delta \theta^i_\text{sys} = \left(\boldsymbol{\Gamma}^{-1}\right)^{ij} \langle \partial_j h | \Delta \mathcal{A} \cos(\Phi) - \mathcal{A} \Delta \Phi \sin(\Phi) \rangle|_{\boldsymbol{\theta}=\boldsymbol{\theta}_\text{A}}, \label{eq:CV29}
% \end{align}
\begin{align}
    \Delta \theta^i_\text{sys} = \left(\boldsymbol{\Gamma}^{-1}\right)^{ij} \langle \partial_j h | \Delta \mathcal{A} \cos(\Phi_\text{A}) - \mathcal{A}_\text{A} \Delta \Phi \sin(\Phi_\text{A}) \rangle|_{\boldsymbol{\theta}=\boldsymbol{\theta}_\text{A}}, \label{eq:CV29}
\end{align}
where $\Delta A = A_0(t; \boldsymbol{\theta}_\text{A}) - A_\text{A}(t; \boldsymbol{\theta}_\text{A}) $ and $\Delta \Phi = \Phi_0(t; \boldsymbol{\theta}_\text{A}) - \Phi_\text{A}(t; \boldsymbol{\theta}_\text{A}) $. 
% \pa{(Are $A$ and $\Phi$ defined here?)} 
Notice that all the amplitudes and phases are evaluated at the parameter values estimated by the approximate waveform. As mentioned in \cite{Cutler_2007}, this formula is valid for $\Delta\theta^i_\text{sys} \Delta\theta^j_\text{sys}\partial_{ij} \Phi \ll 1$. 

\begin{figure}[!tp]
    \centering
    \includegraphics[width = 8.6cm]{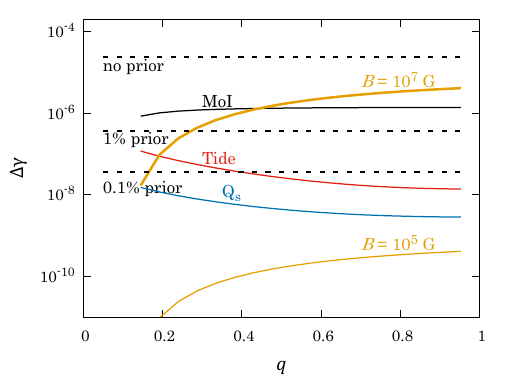}
    \caption{The statistical and systematic errors on $\gamma$ of the DWD system in the first row of Table~\ref{Tab:sources} for $n=-1$ at different mass ratios. Different priors are imposed on the chirp mass when estimating the statistical errors (dashed lines) as described in the right panel of Fig.~\ref{fig:ppE_error}.
    % \ky{How about we show the sys error lines from $B$ also in solid with the same color so that all the sys errors are in solid? For stat errors, can we make the lines thicker? (Same comments apply to Figs.5 and 8.)}
    % \vl{Modified the color and line thickness of Figs. 3,5,8.} \ky{Thanks. How about we make $10^7$G one thick solid and make the same changes to the other figures so that we can have all sys errors as solid?}
    } 
    \label{fig:ppE_error_sys}
\end{figure}

Using the source parameters of ``DWD (sim.)" in Table~\ref{Tab:sources}, we compute systematic errors caused by the astrophysical effects listed in Sec.~\ref{Sec:orbit_evolution} and show the results in Fig.~\ref{fig:ppE_error_sys}. Since the astrophysical effects depend on the individual masses while we only give the chirp mass in the table, we vary the binary's mass ratio, $q$~\footnote{Note that the original simulated source has a fixed mass ratio. Here, we show how the astrophysical effects depend on the mass ratio by varying its value between 0 and 1.}, and fix $n = -1$. 
Notice that the dominant systematic error comes from ignoring moments of inertia or strong magnetic fields, consistent with Fig.~\ref{fig:Delta_astro}.
Without priors on $\mathcal{M}$, the statistical error dominates over the systematic error contribution from the astrophysical effects we consider. To impose better constraints on $\gamma$, priors on $\mathcal{M}$ can be imposed, assuming that we are given independent measurements on the masses of the binary constituents, as described in Sec.~\ref{ssec:statistical_error}. For systems with a large mass ratio, the systematic error becomes significant enough to affect the constraints on $\gamma$ if the chirp mass is independently measured up to 1\% level or below, and therefore should not be ignored in the waveform model.

\subsection{\label{ssec:PE_with_astro} Parameter estimation with astrophysical effects}

As shown above, the astrophysical effects can limit our ability to constrain $\gamma$. In order to place accurate bounds, we need to perform parameter estimation that accounts for these additional effects. In this subsection, we briefly discuss how the statistical error $\Delta \gamma$ gets affected when we include the additional astrophysical parameters to the search parameter set.

The information about the orbital evolution is contained in the frequency evolution of the signal, i.e., within $\dot f_0$ and $\ddot f_0$. In the approximate model, we relate $(\dot f_0, \ddot f_0)$ to $(\mathcal{M}, \gamma)$ through Eqs.~\eqref{eq:df_approx} and \eqref{eq:ddf_approx}. When we include the astrophysical contributions, the waveform model depends on additional parameters like the moment of inertia or the magnetic field. In principle, higher-order derivatives of the frequency need to be measured to determine those. However, that can be challenging given the limited observation time and sensitivity of the detector. We are then prone to having strong degeneracies between the parameters. 

\begin{figure}[!tp]
    \centering
    \includegraphics[width = 8.6cm]{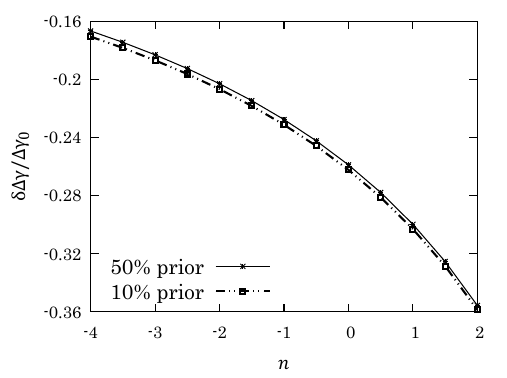}
    \caption{The fractional difference on the statistical error $\Delta \gamma$ with and without the total moment of inertia of the binary in the search parameter set. The error without the moment of inertia in the denominator is denoted by $\Delta \gamma_0$. Gaussian priors with r.m.s. width of either 10\% or 50\% of the fiducial value of the total moment of inertia $I$ is imposed. 
    % \pa{(Give the symbols on the y-axis instead of words ``fractional difference"?)} 
    }
    \label{fig:Fisher_astro}
\end{figure}

On the other hand, some of the astrophysical parameters have scaling relations with the WD masses. We can therefore impose priors on these additional parameters to break the degeneracies. As an example, we consider the statistical error on $\gamma$, denoted by $\Delta \gamma_{I}$, with a  waveform model containing the original 9 parameters and the total moment of inertia of the WDs through a 10$\times$10 Fisher information matrix. For different $n$, the fractional difference $(\Delta \gamma_{I}-\Delta \gamma)/\Delta \gamma$ (where $\Delta\gamma$ is the statistical error on $\gamma$ for 9 parameters without the total moment of inertia) ranges from 15\% to 40\% if we assume a Gaussian prior of 50\% of the fiducial value of $I$ (see Fig.~\ref{fig:Fisher_astro}). The result converges to values slightly smaller than $\Delta \gamma$ as we impose tighter priors. This difference is due to the change in the phase of the signal by including $\Delta_I$, which causes a change in the 
%\vl{Oh, that's right. I checked the values of SNRs and they are equal. I probably made some mistakes before when I tried to explain the consistent difference in $\Delta \gamma$ between the 9- and 10-parameter cases even when I used a very tight prior (hoping it to reduce to 9-parameter case). It should be the Fisher elements having a consistent difference due to the presence of $\Delta_I$ in the phase. I have changed the description below to match what actually happened.} 
Fisher matrix elements. This causes a non-vanishing difference between $\Delta \gamma_{I}$ and $\Delta \gamma$ even if we use an extremely tight prior. Still, the result shows that if we have a certain knowledge of the astrophysical parameters, the constraints on $\gamma$ are expected to be close to the 9-parameter case.

\section{\label{sec:SMG} Application to screened modified gravity}

The SMG is a class of scalar-tensor theories that has a screening mechanism forcing deviations from GR significant only on a large scale, allowing them to pass some of the most stringent constraints from solar system tests while being able to explain cosmological scale observations \cite{Gubser_2004, Zhang_2017, Zhang_2019}. Since the screening mechanism works less efficiently for white dwarfs than neutron stars or black holes, WD binaries may place more stringent bounds on the theory than binary neutron stars or black holes.

The action of a scalar-tensor theory in the Einstein frame is written as \cite{Zhang_2017}
\begin{align}
    S = \int d^4x \sqrt{-g} \left[\frac{M_\text{Pl}^2}{2}R - \frac{1}{2}(\nabla \tilde\Phi)^2 - V(\tilde\Phi)\right] + S_\text{M},
\end{align}
where $M_\text{Pl} = \sqrt{\hbar/(8\pi)}$  is the reduced Planck mass, $V(\tilde\Phi)$ is the bare potential for the scalar field $\tilde \Phi$, and $S_\text{M}$ is the matter action. Note that the matter fields are minimally coupled to the Jordan frame metric, $\bar g_{\alpha\beta}$, which is related to the Einstein frame metric $g_{\alpha\beta}$ through a conformal coupling $\bar g_{\alpha\beta} = A^2(\tilde \Phi) g_{\alpha\beta}$.

The scalar field follows the Klein-Gordon equation
\begin{align}
    \Box \tilde\Phi = \frac{\partial V_\text{eff}}{\partial \tilde \Phi}, 
\end{align}
where $\Box$ is the Einstein frame d'Alembertian, $V_\text{eff}$ is the effective potential that depends on the bare potential and the conformal coupling $A^2(\tilde{\Phi})$ (see \cite{Zhang_2017} for details). The minimum of $V_\text{eff}$ represents the physical vacuum and gives the vacuum expectation value of the field $\Phi_\text{VEV}$. Within SMG theories, this $V_\text{eff}$ has density dependence such that the field around the vacuum acquires an effective mass that increases with density.

\begin{figure}[!tp]
    \centering
    \includegraphics[width = 8.6cm]{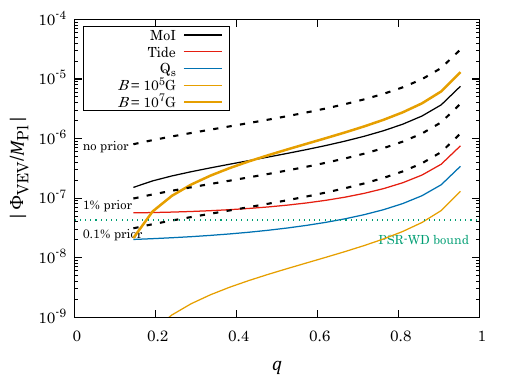}
    \caption{Statistical and systematic errors on the 
    SMG parameter for systems with different mass ratios. Statistical errors with different priors on the chirp mass are shown with dashed lines.
    \label{fig:SMG_bounds}}
\end{figure}

At the leading PN order ($-1$PN), these theories introduce a non-GR effect scaling inversely as the compactness of the stars \cite{Zhang_2017}: 
%\begin{align}
%    \frac{\Phi_\text{VEV}}{M_\text{Pl}} = \sqrt{\frac{192}{5} \gamma \eta^{-2/5} \left(\frac{1}{C_1}-\frac{1}{C_2}\right)^{-2}}, \label{eq:SMG_par}
%\end{align}
\begin{align}
    \gamma = \frac{5}{192} \eta^{2/5} \left(\frac{\Phi_\text{VEV}}{M_\text{Pl}}\right)^2 \left(\frac{1}{C_1}-\frac{1}{C_2}\right)^2 , \label{eq:SMG_par}
\end{align}
where $C_{A} = m_{A}/R_{A}$ is the compactness of the $A$th WD.
Using Eq.~\eqref{eq:SMG_par}, we can convert the measurement errors of $\gamma$  into bounds on the SMG non-GR parameter. We show the results for systems with orbital parameters given in Table~\ref{Tab:sources}, but of various mass ratios, $q = m_1/m_2$, in Fig.~\ref{fig:SMG_bounds}. Due to the dependence on compactness, both the statistical error and systematic error increase with $q$. Similar to the results in Sec.~\ref{ssec:systematic_error}, the statistical error reaches the level of the current observation bounds set by the pulsar-WD binary PSR J1738+0333 \cite{Freire_2012, Zhang_2018} for tight priors on the chirp mass of $\sim$0.1\%. If the DWD system has $q<0.3$, the GW signal can potentially improve the bound on $\Phi_\text{VEV}/M_\text{Pl}$. Even in those optimal cases, the astrophysical effects still need to be accounted for in the waveform model. In Appendix~\ref{app:axion}, we consider the constraints on theories involving axions based on the measurement uncertainties of $\gamma$ and achieve similar constraints as the SMG.

\section{\label{Sec:conclusion} Summary and Conclusion}

In this paper, we study the possibility of testing GR with GW signals from galactic DWD systems using LISA in the presence of various astrophysical effects. We employ the ppE waveform model to parameterize the non-GR effect with $\gamma$ at different PN order $n$. First, we illustrate the ability of  LISA to constrain $\gamma$ with the ppE waveform without the astrophysical effects by computing the Fisher matrix. Compared to GW signals from binary black hole coalescence, the DWD signal can provide more stringent bounds at negative PN order like that of pulsar observations. However, it requires a tight prior on the chirp mass of the system in order to reach the same level as the most stringent binary pulsar bounds.

We then consider the astrophysical effects on the parameter estimation, namely the rotation effect, tidal effect, spin-induced deformation, and dissipation through the unipolar induction within the magnetosphere. Assuming that the waveform model is missing these effects, we estimate the systematic shift of the measured $\gamma$ and show that it becomes significant when the required accuracy of $\gamma$ is near that of the binary pulsar constraint. In other words, one cannot leave out these effects in the waveform model.

Lastly, we apply the results of ppE model to constrain the SMG theory. The non-GR parameter, $\Phi_\text{VEV}/M_\text{Pl}$ has inverse dependence on the compactness of the stars of the binary, making DWDs a good type of source to test this theory. Our results however show that it would be challenging to use DWDs to pose stronger constraints than the current bound from PSR-WD binaries as it requires a tight prior on the chirp mass and accurate modelling of several astrophysical effects involved.

Regarding future work, we have provided some preliminary results showing that the other potential LISA GW sources (see Appendix~\ref{app:other_constraints}) might be useful in testing GR. However, the population of some of these sources is still poorly known. Moreover, many of these sources can still be subjected to astrophysical systematics due to mismodelling. Further studies on the population models and the orbital dynamics are required for such systems. Another important avenue for future work is to confirm our results in full Bayesian analysis with e.g. Markov-chain Monte Carlo or nested sampling method.

\acknowledgments
We all acknowledge support from NASA Grant No. 80NSSC20K0523. 
K.Y. acknowledges support from a Sloan Foundation Research Fellowship and the Owens Family Foundation.

\appendix

\section{\label{app:spinQ} Derivation of the spin-induced quadrupole moment contribution to the frequency evolution}

To derive the correction to the orbital decay rate due to the spin-induced quadrupole moment for a synchronized binary, we first identify the perturbation of the potential energy. Then, we derive the perturbation to the orbital radius as a function of orbital frequency (i.e., modified Kepler's third law). Finally, we apply the energy balance law to obtain the perturbation to the orbital decay rate. For simplicity, we assume the system is spin-aligned and consider the contribution of the quadrupole moment of star 1 only\footnote{The contribution from star 2 can easily be included by taking the correction, changing the index 1 and 2, and adding this to the correction from star 1 only)} and assume that the spin of the star remains synchronized with the orbit. As we sill see, this leads to a slightly different orbital decay rate from \cite{Poisson_1998} that considered binaries without synchronization.

The potential of the binary system is given by
\begin{align}
    V(r) = -\frac{M}{r}\left(1 + \frac{Q_s}{2 m_1 r^2} \right) \label{eq:Qs_potential},
\end{align}
where the spin-induced quadrupole moment scalar is defined through \cite{Laarakkers_1999}
\begin{align}
    Q_{ij} = -Q_s \left(n_i n_j - \frac{1}{3}\delta_{ij}\right),
\end{align}
and the unit vector is set as $\boldsymbol{n} = (0,0,1)$.
The radial component of the equation of motion can be obtained from Eq.~\eqref{eq:Qs_potential}:
\begin{align}
    \ddot{r} - r\Omega^2 = -\frac{M}{r^2}\left(1 + \frac{3Q_s}{2 m_1 r^2} \right),
\end{align}
where $\ddot{r}$ is taken to be zero for circular orbits.
The modified Kepler's law is given by %\ky{Also, you mention in the first paragraph of this section that we'll derive the correction to the orbital radius in terms of the frequency, but the below equation is still in terms of the radius. Can we show the correction to Kepler's law?}
%\ky{I wasn't sure which one you wanted to keep. I like the second line better but then we don't need to define $r_0$.}
\begin{align}
    r = \left(\frac{M}{\Omega^2}\right)^{1/3}\left(1 + \frac{Q_s \Omega^{4/3}}{2 m_1 M^{2/3}}\right).
\end{align}

The change in $\dot f$ is then determined by the rate of energy dissipation
    \begin{align}
        \dot f = \frac{\dot E}{ \pi \frac{dE}{d\Omega}}, \label{eq:df_change}
    \end{align}
    where $\Omega = \dot\phi$ and $\dot E = dE/dt$ is given by
    \begin{align}
        \dot E =& -\frac{32}{5} \mu^2 r^4 \Omega^6, \nonumber\\
        =& -\frac{32}{5}\mu^2 M^{4/3} \Omega^{10/3} \left(1 + 2\frac{Q_s\Omega^{4/3}}{M^{2/3} m_1}\right). \label{eq:dEdt}
    \end{align}
     $\frac{dE}{d\Omega}$ is derived from $E(\Omega)$ as
    \begin{align}
        E(\Omega) =& \frac{1}{2}\mu r^2 \Omega^2 - \frac{M \mu}{r} \left(1 + \frac{Q_s}{2 m_1 r^2}\right) \nonumber\\
        =& -\frac{1}{2}\mu M^{2/3}\Omega^{2/3} \left(1 - \frac{Q_s \Omega^{4/3}}{M^{2/3} m_1}\right)\,.
    \end{align}
    We assume $Q_s \propto \Omega^2$ and $\frac{dE}{d\Omega}$ is then given by
    \begin{align}
        \frac{dE}{d\Omega} = -\frac{1}{3}\mu M^{2/3}\Omega^{-1/3} \left(1 - \frac{6 Q_s \Omega^{4/3}}{M^{2/3} m_1}\right). \label{eq:dEdomega}
    \end{align}
    This leads to a different $\frac{dE}{d\Omega}$ from that in \cite{Poisson_1998} which considered binaries that are not synchronized. 

    By substituting Eqs.~\eqref{eq:dEdt} and \eqref{eq:dEdomega} into Eq.~\eqref{eq:df_change}, we have $\Delta_{Q_s} = \dot f/\dot f_\text{GR} - 1$:
    \begin{align}
        \Delta_{Q_s} = \frac{8Q_s \Omega^{4/3}}{M^{2/3} m_1}.
    \end{align}

\section{\label{Sec:Error_vs_Tobs} Dependence of the statistical error on the observation time}

\begin{figure}
    \includegraphics[width = 8.6cm]{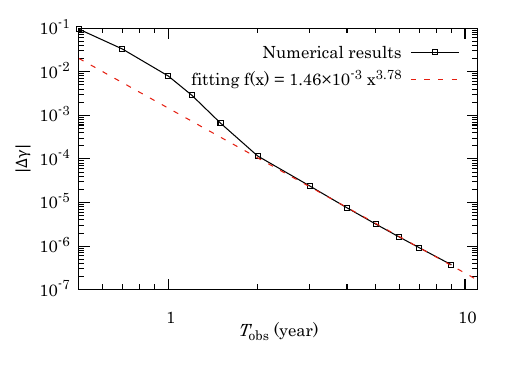}
    \caption{The statistical error on $\gamma$ of the DWD system in Table~\ref{Tab:sources} at different $T_\text{obs}$ for $n=-1$. We also present a fit for the data points with $T_\text{obs} \geq 3~\text{years}$.}
    \label{fig:error_vs_Tobs}
\end{figure}

In this appendix, we study how the statistical error $\Delta \gamma$ scales with the observation time $T_\mathrm{obs}$.
Figure~\ref{fig:error_vs_Tobs} shows the observation time dependence of $\Delta \gamma$ of the DWD system for $n=-1$, which follows a power law with an index of $-3.78$ at large $T_\text{obs}$. Note that this scaling is close to the $T_\text{obs}^{-3.5}$ dependence of the non-GR parameter in Eq.~(17) of \cite{Barbieri_2023}, where they show an approximate expression for the statistical error of the time-variation of the Newtonian gravitational constant using a similar waveform model as the one we use here.

We can also demonstrate this scaling by considering Eq.~\eqref{eq:fisher_td_approx} in the large $T_\text{obs}$ limit. The Fisher matrix element of the ppE parameter is given by
\begin{align}
    \boldsymbol{\Gamma}_{\gamma \gamma} \approx& \frac{\pi^2 \mathcal{A}^2 {x}^{2n}}{38115S_n(f_0)} \bigg[7623 \dot f_\text{GR}^2 T_\text{obs}^5 \nonumber \\
    &+ 385 (11+2n)\dot f_\text{GR} \ddot f_\text{GR} T_\text{obs}^6 +5 \ddot f_\text{GR}^2 (11+2n)^2 T_\text{obs}^7\bigg],
\end{align}
where we have assumed $\mathcal{A}$ is time-independent for simplicity. Hence, for large $T_\text{obs}$, the $T_\text{obs}^7$ term dominates. Similarly,  $\boldsymbol{\Gamma}_{\mathcal{M} \gamma}$ and $\boldsymbol{\Gamma}_{\mathcal{M} \mathcal{M}}$ also have $T_\text{obs}^7$ dependence in this limit. This causes $|\Delta \gamma|$ to scale as $T_\text{obs}^{-3.5}$.

% \vl{Still need to verify this part.}

\section{\label{app:other_constraints} Constraints from other potential LISA sources}

In this section, we repeat the calculation of the statistical uncertainties on $\gamma$ presented in Sec.~\ref{Sec:fisher} and consider its application to SMG as in Sec.~\ref{sec:SMG} for some other potential LISA quasi-monochromatic sources, including neutron star-white dwarf binaries (NS-WD), extreme-mass-ratio-inspiral (XMRI) \cite{Amaro-Seoane_2019}, and the verification binary ZTF J1539+5027 (see Table~\ref{Tab:sources}).

% \vl{Sources that maximize the factors in Eq. (28). It appears to be dominated by the SNR of the source. I did a quick check and found that $M=10^3-10^4\, M_\odot$ EMRIs can give even better constraints if the orbit is close enough. I added a remark in the footnote at the end of this section.} \pa{ sounds good, thanks}

\begin{figure}
    \includegraphics[width = 8.6cm]{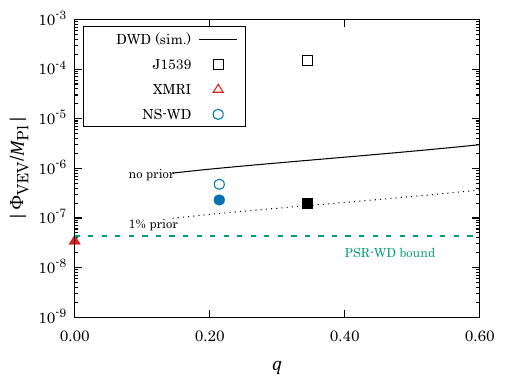}
    \caption{The statistical error on the SMG non-GR parameter against $q$ for various sources that are not considered in Fig.~\ref{fig:SMG_bounds}. The black lines correspond to the statistical error of the source shown in Fig.~\ref{fig:SMG_bounds}. For each source, the open symbol represents the statistical error without prior and the solid symbol represents that with a 1\% Gaussian prior on the chirp mass.}
    \label{fig:SMG_bounds_others}
\end{figure}

In Fig.~\ref{fig:SMG_bounds_others}, we present the statistical uncertainties of the SMG non-GR parameter estimated by the Fisher matrix with the ppE model as described in Sec.~\ref{Sec:fisher} for various LISA sources. For the XMRI source, we assume it is composed of a 0.05$M_\odot$ brown dwarf and a $4\times10^6$$M_\odot$ supermassive black hole, emulating a GW source at the galactic center. All the sources contain at least one star with low compactness and therefore may give strong constraints on the SMG non-GR parameter. From the figure, we see that the statistical errors are comparable but are weaker than the current bound from PSR J1738+0333. One exception is the XMRI, due to its potentially large SNR \cite{Amaro-Seoane_2019} if the orbit is close enough without tidal disruption. The population of these systems is currently uncertain\footnote{In \cite{Amaro-Seoane_2019}, the authors use a steady state power-law distribution function of the brown dwarfs to estimate that there are $\sim$20 such sources within $10^{-3}$~pc of the galactic center of the Milky Way. Out of these, $\sim 5$ are high-frequency sources with circular orbits. However, the actual event rate depends also on the formation and death rate of these systems. Note that some extreme-mass-ratio-inspiral (EMRI) systems consisting of an intermediate-mass black hole within galactic distance can have larger SNRs than the XMRI described here, and therefore can provide even stronger constraints on the non-GR parameters. Due to the lack of constraints on the population and the orbital parameters, we do not include the EMRI results.}.
% \sout{The astrophysical systematics can be smaller for XMRIs than DWDs due to the smaller magnetic field strength of the brown dwarfs.}
% \pa{(BD have radius about 0.1rsun, similar to Jupiter's. By comparison, WD are smaller than BD, not bigger. A typical WD radius might be 0.01rsun. Their observed magnetic fields are smaller though, more at the kG level while WD can be up to MG-GG.}. 
% \vl{However, the astrophysical systematics can be large due to the larger radii of the brown dwarfs compared to WDs, even though those from magnetic interactions are expected to be lower due to the smaller magnetic field strength. }
This again means the GW measurement alone is insufficient to improve the bound on the SMG theories. Prior information on the chirp mass (e.g., 1\% Gaussian priors on the chirp mass, as indicated by the solid symbols in Fig.~\ref{fig:SMG_bounds_others}) is required for stronger constraints. 

\section{\label{app:axion} Constraining axion-like particles}

% \ky{I think the title of this section should be more specific, like Constraining axion-like particles.}

% \ky{We need to cite Appendices C and D somewhere and change the order of the appendix sections if necessary.}

Theories involving axions are also a good candidate to test with DWD systems as the axion charges become larger for less compact stars \cite{Hook:2017psm,Huang:2018pbu,Huang_2019}. Axion-like particles (ALPs) are pseudo-scalar fields that extend the standard model. One example is the QCD axions that are introduced to resolve the strong CP problem in quantum chromodynamics (QCD) \cite{Peccei_1977, Weinberg_1978, Wilczek_1978}. ALPs are also popular dark matter (DM) candidates \cite{Marsh_2016}. For QCD axions, it has been shown to account for the observed DM abundance if the decay constant is above $10^{12}$GeV \cite{Preskill_1983}. The ALP parameters have been constrained by both laboratory experiments \cite{Arik_2011, Asztalos_2004, Budker_2014} and astrophysical observations \cite{Raffelt_1986, Ellis_1987, Janka_1996}. The addition of GW observations by LISA can provide an independent probe of such particles.

During inspiral, the extra force from the ALPs affects the orbital phase. In particular, the scalar Larmor radiation causes a change in the orbital decay rate \cite{Huang_2019, Poddar_2020}. The effect depends on the axion dipole moment $p$ of a binary sourced by the axion charges, which can be approximated by 
\begin{align}
    p = 4\pi f_a \mathcal{M} \eta^{2/5} r \left(C_1^{-1}-C_2^{-1}\right),
\end{align}
where $r$ is the orbital separation, $f_a$ is the axion decay constant (see \cite{Poddar_2020}). 
% \ky{What is $r$? orbital separation?} 
Relating to the ppE parameter, $\gamma$, defined in Eqs.~\eqref{eq:df_approx} and \eqref{eq:ddf_approx}, we have
%\begin{align}
%f_a = \sqrt{\frac{48}{5\pi} \frac{ \hbar c^5}{G} \eta^{-2/5} \gamma \left(C_1^{-1}-C_2^{-1}\right)^{-2}}. \label{eq:axion_par}
%\end{align}
\begin{align}
\gamma = \frac{5\pi}{48}\eta^{2/5} \frac{G}{\hbar c^5} f_a^2 \left(\frac{1}{C_1}-\frac{1}{C_2}\right)^2. \label{eq:axion_par}
\end{align}
Notice that Eq.~\eqref{eq:axion_par} has similar dependence on $\eta$, 
 $\gamma$ and compactness as the SMG parameter in Eq.~\eqref{eq:SMG_par}. Hence, similar constraints can be found using the DWD systems as shown in Fig.~\ref{fig:SMG_bounds}. 

\begin{figure}
    \includegraphics[width = 8.6cm]{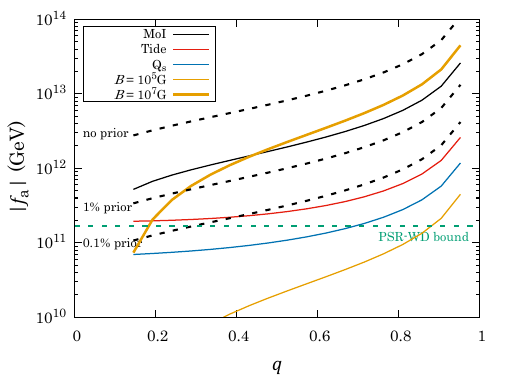}
    \caption{Similar to Fig.~\ref{fig:SMG_bounds} but showing the statistical and systematic errors of the axion decay constant $f_a$ against the mass ratio $q$. }
    \label{fig:axion_bounds}
\end{figure}

In Fig.~\ref{fig:axion_bounds}, we show the statistical and systematic error of $f_a$ in the same manner as Fig.~\ref{fig:SMG_bounds}. The statistical errors are obtained by imposing a Gaussian prior of different widths on the chirp mass and the current bound is obtained from the pulsar-WD binary PSR J0348+0432 \cite{Antoniadis_2013} (a similar but less stringent bound is obtained from J1738+0333). Due to the similarity of the dependence of the non-GR parameter on the WD parameters, the constraints obtained are qualitatively the same as SMG. 

\nocite{*}

\bibliography{apssamp}% Produces the bibliography via BibTeX.

%merlin.mbs apsrev4-1.bst 2010-07-25 4.21a (PWD, AO, DPC) hacked
%Control: key (0)
%Control: author (8) initials jnrlst
%Control: editor formatted (1) identically to author
%Control: production of article title (-1) disabled
%Control: page (0) single
%Control: year (1) truncated
%Control: production of eprint (0) enabled
\begin{thebibliography}{108}%
\makeatletter
\providecommand \@ifxundefined [1]{%
 \@ifx{#1\undefined}
}%
\providecommand \@ifnum [1]{%
 \ifnum #1\expandafter \@firstoftwo
 \else \expandafter \@secondoftwo
 \fi
}%
\providecommand \@ifx [1]{%
 \ifx #1\expandafter \@firstoftwo
 \else \expandafter \@secondoftwo
 \fi
}%
\providecommand \natexlab [1]{#1}%
\providecommand \enquote  [1]{``#1''}%
\providecommand \bibnamefont  [1]{#1}%
\providecommand \bibfnamefont [1]{#1}%
\providecommand \citenamefont [1]{#1}%
\providecommand \href@noop [0]{\@secondoftwo}%
\providecommand \href [0]{\begingroup \@sanitize@url \@href}%
\providecommand \@href[1]{\@@startlink{#1}\@@href}%
\providecommand \@@href[1]{\endgroup#1\@@endlink}%
\providecommand \@sanitize@url [0]{\catcode `\\12\catcode `\$12\catcode
  `\&12\catcode `\#12\catcode `\^12\catcode `\_12\catcode `\%12\relax}%
\providecommand \@@startlink[1]{}%
\providecommand \@@endlink[0]{}%
\providecommand \url  [0]{\begingroup\@sanitize@url \@url }%
\providecommand \@url [1]{\endgroup\@href {#1}{\urlprefix }}%
\providecommand \urlprefix  [0]{URL }%
\providecommand \Eprint [0]{\href }%
\providecommand \doibase [0]{http://dx.doi.org/}%
\providecommand \selectlanguage [0]{\@gobble}%
\providecommand \bibinfo  [0]{\@secondoftwo}%
\providecommand \bibfield  [0]{\@secondoftwo}%
\providecommand \translation [1]{[#1]}%
\providecommand \BibitemOpen [0]{}%
\providecommand \bibitemStop [0]{}%
\providecommand \bibitemNoStop [0]{.\EOS\space}%
\providecommand \EOS [0]{\spacefactor3000\relax}%
\providecommand \BibitemShut  [1]{\csname bibitem#1\endcsname}%
\let\auto@bib@innerbib\@empty
%</preamble>
\bibitem [{\citenamefont {Brans}\ and\ \citenamefont
  {Dicke}(1961)}]{Brans_Dicke_1961}%
  \BibitemOpen
  \bibfield  {author} {\bibinfo {author} {\bibfnamefont {C.}~\bibnamefont
  {Brans}}\ and\ \bibinfo {author} {\bibfnamefont {R.~H.}\ \bibnamefont
  {Dicke}},\ }\href {\doibase 10.1103/PhysRev.124.925} {\bibfield  {journal}
  {\bibinfo  {journal} {Phys. Rev.}\ }\textbf {\bibinfo {volume} {124}},\
  \bibinfo {pages} {925} (\bibinfo {year} {1961})}\BibitemShut {NoStop}%
\bibitem [{\citenamefont {Damour}\ and\ \citenamefont
  {Esposito-Farese}(1992)}]{Damour_1992}%
  \BibitemOpen
  \bibfield  {author} {\bibinfo {author} {\bibfnamefont {T.}~\bibnamefont
  {Damour}}\ and\ \bibinfo {author} {\bibfnamefont {G.}~\bibnamefont
  {Esposito-Farese}},\ }\href {\doibase 10.1088/0264-9381/9/9/015} {\bibfield
  {journal} {\bibinfo  {journal} {Class. Quantum Gravity}\ }\textbf {\bibinfo
  {volume} {9}},\ \bibinfo {pages} {2093} (\bibinfo {year} {1992})}\BibitemShut
  {NoStop}%
\bibitem [{\citenamefont {Barausse}\ \emph {et~al.}(2013)\citenamefont
  {Barausse}, \citenamefont {Palenzuela}, \citenamefont {Ponce},\ and\
  \citenamefont {Lehner}}]{PhysRevD.87.081506}%
  \BibitemOpen
  \bibfield  {author} {\bibinfo {author} {\bibfnamefont {E.}~\bibnamefont
  {Barausse}}, \bibinfo {author} {\bibfnamefont {C.}~\bibnamefont
  {Palenzuela}}, \bibinfo {author} {\bibfnamefont {M.}~\bibnamefont {Ponce}}, \
  and\ \bibinfo {author} {\bibfnamefont {L.}~\bibnamefont {Lehner}},\ }\href
  {\doibase 10.1103/PhysRevD.87.081506} {\bibfield  {journal} {\bibinfo
  {journal} {Phys. Rev. D}\ }\textbf {\bibinfo {volume} {87}},\ \bibinfo
  {pages} {081506} (\bibinfo {year} {2013})}\BibitemShut {NoStop}%
\bibitem [{\citenamefont {De~Felice}\ and\ \citenamefont
  {Tsujikawa}(2010)}]{DeFelice:2010aj}%
  \BibitemOpen
  \bibfield  {author} {\bibinfo {author} {\bibfnamefont {A.}~\bibnamefont
  {De~Felice}}\ and\ \bibinfo {author} {\bibfnamefont {S.}~\bibnamefont
  {Tsujikawa}},\ }\href {\doibase 10.12942/lrr-2010-3} {\bibfield  {journal}
  {\bibinfo  {journal} {Living Rev. Rel.}\ }\textbf {\bibinfo {volume} {13}},\
  \bibinfo {pages} {3} (\bibinfo {year} {2010})},\ \Eprint
  {http://arxiv.org/abs/1002.4928} {arXiv:1002.4928 [gr-qc]} \BibitemShut
  {NoStop}%
\bibitem [{\citenamefont {Horndeski}(1974)}]{Horndeski:1974wa}%
  \BibitemOpen
  \bibfield  {author} {\bibinfo {author} {\bibfnamefont {G.~W.}\ \bibnamefont
  {Horndeski}},\ }\href {\doibase 10.1007/BF01807638} {\bibfield  {journal}
  {\bibinfo  {journal} {Int. J. Theor. Phys.}\ }\textbf {\bibinfo {volume}
  {10}},\ \bibinfo {pages} {363} (\bibinfo {year} {1974})}\BibitemShut
  {NoStop}%
\bibitem [{\citenamefont {Kobayashi}(2019)}]{Kobayashi:2019hrl}%
  \BibitemOpen
  \bibfield  {author} {\bibinfo {author} {\bibfnamefont {T.}~\bibnamefont
  {Kobayashi}},\ }\href {\doibase 10.1088/1361-6633/ab2429} {\bibfield
  {journal} {\bibinfo  {journal} {Rept. Prog. Phys.}\ }\textbf {\bibinfo
  {volume} {82}},\ \bibinfo {pages} {086901} (\bibinfo {year} {2019})},\
  \Eprint {http://arxiv.org/abs/1901.07183} {arXiv:1901.07183 [gr-qc]}
  \BibitemShut {NoStop}%
\bibitem [{\citenamefont {Brax}\ \emph {et~al.}(2012)\citenamefont {Brax},
  \citenamefont {Davis}, \citenamefont {Li},\ and\ \citenamefont
  {Winther}}]{Brax_2012}%
  \BibitemOpen
  \bibfield  {author} {\bibinfo {author} {\bibfnamefont {P.}~\bibnamefont
  {Brax}}, \bibinfo {author} {\bibfnamefont {A.-C.}\ \bibnamefont {Davis}},
  \bibinfo {author} {\bibfnamefont {B.}~\bibnamefont {Li}}, \ and\ \bibinfo
  {author} {\bibfnamefont {H.~A.}\ \bibnamefont {Winther}},\ }\href {\doibase
  10.1103/PhysRevD.86.044015} {\bibfield  {journal} {\bibinfo  {journal} {Phys.
  Rev. D}\ }\textbf {\bibinfo {volume} {86}},\ \bibinfo {pages} {044015}
  (\bibinfo {year} {2012})}\BibitemShut {NoStop}%
\bibitem [{\citenamefont {Gubser}\ and\ \citenamefont
  {Khoury}(2004)}]{Gubser_2004}%
  \BibitemOpen
  \bibfield  {author} {\bibinfo {author} {\bibfnamefont {S.~S.}\ \bibnamefont
  {Gubser}}\ and\ \bibinfo {author} {\bibfnamefont {J.}~\bibnamefont
  {Khoury}},\ }\href {\doibase 10.1103/PhysRevD.70.104001} {\bibfield
  {journal} {\bibinfo  {journal} {Phys. Rev. D}\ }\textbf {\bibinfo {volume}
  {70}},\ \bibinfo {pages} {104001} (\bibinfo {year} {2004})}\BibitemShut
  {NoStop}%
\bibitem [{\citenamefont {Jacobson}\ and\ \citenamefont
  {Mattingly}(2001)}]{Jacobson_2001}%
  \BibitemOpen
  \bibfield  {author} {\bibinfo {author} {\bibfnamefont {T.}~\bibnamefont
  {Jacobson}}\ and\ \bibinfo {author} {\bibfnamefont {D.}~\bibnamefont
  {Mattingly}},\ }\href {\doibase 10.1103/PhysRevD.64.024028} {\bibfield
  {journal} {\bibinfo  {journal} {Phys. Rev. D}\ }\textbf {\bibinfo {volume}
  {64}},\ \bibinfo {pages} {024028} (\bibinfo {year} {2001})}\BibitemShut
  {NoStop}%
\bibitem [{\citenamefont {Jacobson}(2007)}]{Jacobson:2007veq}%
  \BibitemOpen
  \bibfield  {author} {\bibinfo {author} {\bibfnamefont {T.}~\bibnamefont
  {Jacobson}},\ }\href {\doibase 10.22323/1.043.0020} {\bibfield  {journal}
  {\bibinfo  {journal} {PoS}\ }\textbf {\bibinfo {volume} {QG-PH}},\ \bibinfo
  {pages} {020} (\bibinfo {year} {2007})},\ \Eprint
  {http://arxiv.org/abs/0801.1547} {arXiv:0801.1547 [gr-qc]} \BibitemShut
  {NoStop}%
\bibitem [{\citenamefont {Blas}\ \emph {et~al.}(2011)\citenamefont {Blas},
  \citenamefont {Pujolas},\ and\ \citenamefont {Sibiryakov}}]{Blas:2010hb}%
  \BibitemOpen
  \bibfield  {author} {\bibinfo {author} {\bibfnamefont {D.}~\bibnamefont
  {Blas}}, \bibinfo {author} {\bibfnamefont {O.}~\bibnamefont {Pujolas}}, \
  and\ \bibinfo {author} {\bibfnamefont {S.}~\bibnamefont {Sibiryakov}},\
  }\href {\doibase 10.1007/JHEP04(2011)018} {\bibfield  {journal} {\bibinfo
  {journal} {JHEP}\ }\textbf {\bibinfo {volume} {04}},\ \bibinfo {pages} {018}
  (\bibinfo {year} {2011})},\ \Eprint {http://arxiv.org/abs/1007.3503}
  {arXiv:1007.3503 [hep-th]} \BibitemShut {NoStop}%
\bibitem [{\citenamefont {Kanti}\ \emph {et~al.}(1996)\citenamefont {Kanti},
  \citenamefont {Mavromatos}, \citenamefont {Rizos}, \citenamefont {Tamvakis},\
  and\ \citenamefont {Winstanley}}]{Kanti_1996}%
  \BibitemOpen
  \bibfield  {author} {\bibinfo {author} {\bibfnamefont {P.}~\bibnamefont
  {Kanti}}, \bibinfo {author} {\bibfnamefont {N.~E.}\ \bibnamefont
  {Mavromatos}}, \bibinfo {author} {\bibfnamefont {J.}~\bibnamefont {Rizos}},
  \bibinfo {author} {\bibfnamefont {K.}~\bibnamefont {Tamvakis}}, \ and\
  \bibinfo {author} {\bibfnamefont {E.}~\bibnamefont {Winstanley}},\ }\href
  {\doibase 10.1103/PhysRevD.54.5049} {\bibfield  {journal} {\bibinfo
  {journal} {Phys. Rev. D}\ }\textbf {\bibinfo {volume} {54}},\ \bibinfo
  {pages} {5049} (\bibinfo {year} {1996})}\BibitemShut {NoStop}%
\bibitem [{\citenamefont {Jackiw}\ and\ \citenamefont
  {Pi}(2003)}]{Jackiw_2003}%
  \BibitemOpen
  \bibfield  {author} {\bibinfo {author} {\bibfnamefont {R.}~\bibnamefont
  {Jackiw}}\ and\ \bibinfo {author} {\bibfnamefont {S.-Y.}\ \bibnamefont
  {Pi}},\ }\href {\doibase 10.1103/PhysRevD.68.104012} {\bibfield  {journal}
  {\bibinfo  {journal} {Phys. Rev. D}\ }\textbf {\bibinfo {volume} {68}},\
  \bibinfo {pages} {104012} (\bibinfo {year} {2003})}\BibitemShut {NoStop}%
\bibitem [{\citenamefont {Alexander}\ and\ \citenamefont
  {Yunes}(2009)}]{Alexander_2009}%
  \BibitemOpen
  \bibfield  {author} {\bibinfo {author} {\bibfnamefont {S.}~\bibnamefont
  {Alexander}}\ and\ \bibinfo {author} {\bibfnamefont {N.}~\bibnamefont
  {Yunes}},\ }\href {\doibase https://doi.org/10.1016/j.physrep.2009.07.002}
  {\bibfield  {journal} {\bibinfo  {journal} {Phys. Rep.}\ }\textbf {\bibinfo
  {volume} {480}},\ \bibinfo {pages} {1} (\bibinfo {year} {2009})}\BibitemShut
  {NoStop}%
\bibitem [{\citenamefont {{Yunes}}\ and\ \citenamefont
  {{Siemens}}(2013)}]{Yunes_2013_rev}%
  \BibitemOpen
  \bibfield  {author} {\bibinfo {author} {\bibfnamefont {N.}~\bibnamefont
  {{Yunes}}}\ and\ \bibinfo {author} {\bibfnamefont {X.}~\bibnamefont
  {{Siemens}}},\ }\href {\doibase 10.12942/lrr-2013-9} {\bibfield  {journal}
  {\bibinfo  {journal} {Living Rev. Relativ.}\ }\textbf {\bibinfo {volume}
  {16}},\ \bibinfo {eid} {9} (\bibinfo {year} {2013})},\ \Eprint
  {http://arxiv.org/abs/1304.3473} {arXiv:1304.3473 [gr-qc]} \BibitemShut
  {NoStop}%
\bibitem [{\citenamefont {Yunes}\ \emph {et~al.}(2024)\citenamefont {Yunes},
  \citenamefont {Siemens},\ and\ \citenamefont {Yagi}}]{Yunes:2024lzm}%
  \BibitemOpen
  \bibfield  {author} {\bibinfo {author} {\bibfnamefont {N.}~\bibnamefont
  {Yunes}}, \bibinfo {author} {\bibfnamefont {X.}~\bibnamefont {Siemens}}, \
  and\ \bibinfo {author} {\bibfnamefont {K.}~\bibnamefont {Yagi}},\ }\href@noop
  {} {\  (\bibinfo {year} {2024})},\ \Eprint {http://arxiv.org/abs/2408.05240}
  {arXiv:2408.05240 [gr-qc]} \BibitemShut {NoStop}%
\bibitem [{\citenamefont {Berti}\ \emph {et~al.}(2015)\citenamefont {Berti}
  \emph {et~al.}}]{Berti:2015itd}%
  \BibitemOpen
  \bibfield  {author} {\bibinfo {author} {\bibfnamefont {E.}~\bibnamefont
  {Berti}} \emph {et~al.},\ }\href {\doibase 10.1088/0264-9381/32/24/243001}
  {\bibfield  {journal} {\bibinfo  {journal} {Class. Quant. Grav.}\ }\textbf
  {\bibinfo {volume} {32}},\ \bibinfo {pages} {243001} (\bibinfo {year}
  {2015})},\ \Eprint {http://arxiv.org/abs/1501.07274} {arXiv:1501.07274
  [gr-qc]} \BibitemShut {NoStop}%
\bibitem [{\citenamefont {Will}(2018)}]{Will_2018}%
  \BibitemOpen
  \bibfield  {author} {\bibinfo {author} {\bibfnamefont {C.~M.}\ \bibnamefont
  {Will}},\ }\href@noop {} {\emph {\bibinfo {title} {{Theory and Experiment in
  Gravitational Physics}}}}\ (\bibinfo  {publisher} {Cambridge University
  Press},\ \bibinfo {year} {2018})\BibitemShut {NoStop}%
\bibitem [{\citenamefont {Abbott}\ \emph {et~al.}(2016)\citenamefont {Abbott}
  \emph {et~al.}}]{LIGOScientific:2016lio}%
  \BibitemOpen
  \bibfield  {author} {\bibinfo {author} {\bibfnamefont {B.~P.}\ \bibnamefont
  {Abbott}} \emph {et~al.} (\bibinfo {collaboration} {LIGO Scientific,
  Virgo}),\ }\href {\doibase 10.1103/PhysRevLett.116.221101} {\bibfield
  {journal} {\bibinfo  {journal} {Phys. Rev. Lett.}\ }\textbf {\bibinfo
  {volume} {116}},\ \bibinfo {pages} {221101} (\bibinfo {year} {2016})},\
  \bibinfo {note} {[Erratum: Phys.Rev.Lett. 121, 129902 (2018)]},\ \Eprint
  {http://arxiv.org/abs/1602.03841} {arXiv:1602.03841 [gr-qc]} \BibitemShut
  {NoStop}%
\bibitem [{\citenamefont {Yunes}\ \emph
  {et~al.}(2016{\natexlab{a}})\citenamefont {Yunes}, \citenamefont {Yagi},\
  and\ \citenamefont {Pretorius}}]{Yunes:2016jcc}%
  \BibitemOpen
  \bibfield  {author} {\bibinfo {author} {\bibfnamefont {N.}~\bibnamefont
  {Yunes}}, \bibinfo {author} {\bibfnamefont {K.}~\bibnamefont {Yagi}}, \ and\
  \bibinfo {author} {\bibfnamefont {F.}~\bibnamefont {Pretorius}},\ }\href
  {\doibase 10.1103/PhysRevD.94.084002} {\bibfield  {journal} {\bibinfo
  {journal} {Phys. Rev. D}\ }\textbf {\bibinfo {volume} {94}},\ \bibinfo
  {pages} {084002} (\bibinfo {year} {2016}{\natexlab{a}})},\ \Eprint
  {http://arxiv.org/abs/1603.08955} {arXiv:1603.08955 [gr-qc]} \BibitemShut
  {NoStop}%
\bibitem [{\citenamefont {Berti}\ \emph
  {et~al.}(2018{\natexlab{a}})\citenamefont {Berti}, \citenamefont {Yagi},\
  and\ \citenamefont {Yunes}}]{Berti:2018cxi}%
  \BibitemOpen
  \bibfield  {author} {\bibinfo {author} {\bibfnamefont {E.}~\bibnamefont
  {Berti}}, \bibinfo {author} {\bibfnamefont {K.}~\bibnamefont {Yagi}}, \ and\
  \bibinfo {author} {\bibfnamefont {N.}~\bibnamefont {Yunes}},\ }\href
  {\doibase 10.1007/s10714-018-2362-8} {\bibfield  {journal} {\bibinfo
  {journal} {Gen. Rel. Grav.}\ }\textbf {\bibinfo {volume} {50}},\ \bibinfo
  {pages} {46} (\bibinfo {year} {2018}{\natexlab{a}})},\ \Eprint
  {http://arxiv.org/abs/1801.03208} {arXiv:1801.03208 [gr-qc]} \BibitemShut
  {NoStop}%
\bibitem [{\citenamefont {Berti}\ \emph
  {et~al.}(2018{\natexlab{b}})\citenamefont {Berti}, \citenamefont {Yagi},
  \citenamefont {Yang},\ and\ \citenamefont {Yunes}}]{Berti:2018vdi}%
  \BibitemOpen
  \bibfield  {author} {\bibinfo {author} {\bibfnamefont {E.}~\bibnamefont
  {Berti}}, \bibinfo {author} {\bibfnamefont {K.}~\bibnamefont {Yagi}},
  \bibinfo {author} {\bibfnamefont {H.}~\bibnamefont {Yang}}, \ and\ \bibinfo
  {author} {\bibfnamefont {N.}~\bibnamefont {Yunes}},\ }\href {\doibase
  10.1007/s10714-018-2372-6} {\bibfield  {journal} {\bibinfo  {journal} {Gen.
  Rel. Grav.}\ }\textbf {\bibinfo {volume} {50}},\ \bibinfo {pages} {49}
  (\bibinfo {year} {2018}{\natexlab{b}})},\ \Eprint
  {http://arxiv.org/abs/1801.03587} {arXiv:1801.03587 [gr-qc]} \BibitemShut
  {NoStop}%
\bibitem [{\citenamefont {Abbott}\ \emph {et~al.}(2021)\citenamefont {Abbott}
  \emph {et~al.}}]{LIGOScientific:2021sio}%
  \BibitemOpen
  \bibfield  {author} {\bibinfo {author} {\bibfnamefont {R.}~\bibnamefont
  {Abbott}} \emph {et~al.} (\bibinfo {collaboration} {LIGO Scientific, VIRGO,
  KAGRA}),\ }\href@noop {} {\  (\bibinfo {year} {2021})},\ \Eprint
  {http://arxiv.org/abs/2112.06861} {arXiv:2112.06861 [gr-qc]} \BibitemShut
  {NoStop}%
\bibitem [{\citenamefont {Abbott}\ \emph {et~al.}(2019)\citenamefont {Abbott}
  \emph {et~al.}}]{LIGOScientific:2018dkp}%
  \BibitemOpen
  \bibfield  {author} {\bibinfo {author} {\bibfnamefont {B.~P.}\ \bibnamefont
  {Abbott}} \emph {et~al.} (\bibinfo {collaboration} {LIGO Scientific,
  Virgo}),\ }\href {\doibase 10.1103/PhysRevLett.123.011102} {\bibfield
  {journal} {\bibinfo  {journal} {Phys. Rev. Lett.}\ }\textbf {\bibinfo
  {volume} {123}},\ \bibinfo {pages} {011102} (\bibinfo {year} {2019})},\
  \Eprint {http://arxiv.org/abs/1811.00364} {arXiv:1811.00364 [gr-qc]}
  \BibitemShut {NoStop}%
\bibitem [{\citenamefont {Takeda}\ \emph {et~al.}(2022)\citenamefont {Takeda},
  \citenamefont {Morisaki},\ and\ \citenamefont {Nishizawa}}]{Takeda:2021hgo}%
  \BibitemOpen
  \bibfield  {author} {\bibinfo {author} {\bibfnamefont {H.}~\bibnamefont
  {Takeda}}, \bibinfo {author} {\bibfnamefont {S.}~\bibnamefont {Morisaki}}, \
  and\ \bibinfo {author} {\bibfnamefont {A.}~\bibnamefont {Nishizawa}},\ }\href
  {\doibase 10.1103/PhysRevD.105.084019} {\bibfield  {journal} {\bibinfo
  {journal} {Phys. Rev. D}\ }\textbf {\bibinfo {volume} {105}},\ \bibinfo
  {pages} {084019} (\bibinfo {year} {2022})},\ \Eprint
  {http://arxiv.org/abs/2105.00253} {arXiv:2105.00253 [gr-qc]} \BibitemShut
  {NoStop}%
\bibitem [{\citenamefont {{Amaro-Seoane}}\ \emph {et~al.}()\citenamefont
  {{Amaro-Seoane}} \emph {et~al.}}]{2017arXiv170200786A}%
  \BibitemOpen
  \bibfield  {author} {\bibinfo {author} {\bibfnamefont {P.}~\bibnamefont
  {{Amaro-Seoane}}} \emph {et~al.},\ }\href@noop {} {}\Eprint
  {http://arxiv.org/abs/1702.00786} {arXiv:1702.00786} \BibitemShut {NoStop}%
\bibitem [{\citenamefont {Perkins}\ \emph {et~al.}(2021)\citenamefont
  {Perkins}, \citenamefont {Yunes},\ and\ \citenamefont
  {Berti}}]{Perkins:2020tra}%
  \BibitemOpen
  \bibfield  {author} {\bibinfo {author} {\bibfnamefont {S.~E.}\ \bibnamefont
  {Perkins}}, \bibinfo {author} {\bibfnamefont {N.}~\bibnamefont {Yunes}}, \
  and\ \bibinfo {author} {\bibfnamefont {E.}~\bibnamefont {Berti}},\ }\href
  {\doibase 10.1103/PhysRevD.103.044024} {\bibfield  {journal} {\bibinfo
  {journal} {Phys. Rev. D}\ }\textbf {\bibinfo {volume} {103}},\ \bibinfo
  {pages} {044024} (\bibinfo {year} {2021})},\ \Eprint
  {http://arxiv.org/abs/2010.09010} {arXiv:2010.09010 [gr-qc]} \BibitemShut
  {NoStop}%
\bibitem [{\citenamefont {Berti}\ \emph {et~al.}(2005)\citenamefont {Berti},
  \citenamefont {Buonanno},\ and\ \citenamefont {Will}}]{Berti:2004bd}%
  \BibitemOpen
  \bibfield  {author} {\bibinfo {author} {\bibfnamefont {E.}~\bibnamefont
  {Berti}}, \bibinfo {author} {\bibfnamefont {A.}~\bibnamefont {Buonanno}}, \
  and\ \bibinfo {author} {\bibfnamefont {C.~M.}\ \bibnamefont {Will}},\ }\href
  {\doibase 10.1103/PhysRevD.71.084025} {\bibfield  {journal} {\bibinfo
  {journal} {Phys. Rev. D}\ }\textbf {\bibinfo {volume} {71}},\ \bibinfo
  {pages} {084025} (\bibinfo {year} {2005})},\ \Eprint
  {http://arxiv.org/abs/gr-qc/0411129} {arXiv:gr-qc/0411129} \BibitemShut
  {NoStop}%
\bibitem [{\citenamefont {Yagi}\ and\ \citenamefont
  {Tanaka}(2010)}]{Yagi:2009zm}%
  \BibitemOpen
  \bibfield  {author} {\bibinfo {author} {\bibfnamefont {K.}~\bibnamefont
  {Yagi}}\ and\ \bibinfo {author} {\bibfnamefont {T.}~\bibnamefont {Tanaka}},\
  }\href {\doibase 10.1103/PhysRevD.81.109902} {\bibfield  {journal} {\bibinfo
  {journal} {Phys. Rev. D}\ }\textbf {\bibinfo {volume} {81}},\ \bibinfo
  {pages} {064008} (\bibinfo {year} {2010})},\ \bibinfo {note} {[Erratum:
  Phys.Rev.D 81, 109902 (2010)]},\ \Eprint {http://arxiv.org/abs/0906.4269}
  {arXiv:0906.4269 [gr-qc]} \BibitemShut {NoStop}%
\bibitem [{\citenamefont {Berti}\ \emph {et~al.}(2011)\citenamefont {Berti},
  \citenamefont {Gair},\ and\ \citenamefont {Sesana}}]{Berti:2011jz}%
  \BibitemOpen
  \bibfield  {author} {\bibinfo {author} {\bibfnamefont {E.}~\bibnamefont
  {Berti}}, \bibinfo {author} {\bibfnamefont {J.}~\bibnamefont {Gair}}, \ and\
  \bibinfo {author} {\bibfnamefont {A.}~\bibnamefont {Sesana}},\ }\href
  {\doibase 10.1103/PhysRevD.84.101501} {\bibfield  {journal} {\bibinfo
  {journal} {Phys. Rev. D}\ }\textbf {\bibinfo {volume} {84}},\ \bibinfo
  {pages} {101501} (\bibinfo {year} {2011})},\ \Eprint
  {http://arxiv.org/abs/1107.3528} {arXiv:1107.3528 [gr-qc]} \BibitemShut
  {NoStop}%
\bibitem [{\citenamefont {Ryan}(1997)}]{Ryan:1997hg}%
  \BibitemOpen
  \bibfield  {author} {\bibinfo {author} {\bibfnamefont {F.~D.}\ \bibnamefont
  {Ryan}},\ }\href {\doibase 10.1103/PhysRevD.56.1845} {\bibfield  {journal}
  {\bibinfo  {journal} {Phys. Rev. D}\ }\textbf {\bibinfo {volume} {56}},\
  \bibinfo {pages} {1845} (\bibinfo {year} {1997})}\BibitemShut {NoStop}%
\bibitem [{\citenamefont {Berti}\ \emph {et~al.}(2006)\citenamefont {Berti},
  \citenamefont {Cardoso},\ and\ \citenamefont {Will}}]{Berti:2005ys}%
  \BibitemOpen
  \bibfield  {author} {\bibinfo {author} {\bibfnamefont {E.}~\bibnamefont
  {Berti}}, \bibinfo {author} {\bibfnamefont {V.}~\bibnamefont {Cardoso}}, \
  and\ \bibinfo {author} {\bibfnamefont {C.~M.}\ \bibnamefont {Will}},\ }\href
  {\doibase 10.1103/PhysRevD.73.064030} {\bibfield  {journal} {\bibinfo
  {journal} {Phys. Rev. D}\ }\textbf {\bibinfo {volume} {73}},\ \bibinfo
  {pages} {064030} (\bibinfo {year} {2006})},\ \Eprint
  {http://arxiv.org/abs/gr-qc/0512160} {arXiv:gr-qc/0512160} \BibitemShut
  {NoStop}%
\bibitem [{\citenamefont {Berti}\ \emph {et~al.}(2016)\citenamefont {Berti},
  \citenamefont {Sesana}, \citenamefont {Barausse}, \citenamefont {Cardoso},\
  and\ \citenamefont {Belczynski}}]{Berti:2016lat}%
  \BibitemOpen
  \bibfield  {author} {\bibinfo {author} {\bibfnamefont {E.}~\bibnamefont
  {Berti}}, \bibinfo {author} {\bibfnamefont {A.}~\bibnamefont {Sesana}},
  \bibinfo {author} {\bibfnamefont {E.}~\bibnamefont {Barausse}}, \bibinfo
  {author} {\bibfnamefont {V.}~\bibnamefont {Cardoso}}, \ and\ \bibinfo
  {author} {\bibfnamefont {K.}~\bibnamefont {Belczynski}},\ }\href {\doibase
  10.1103/PhysRevLett.117.101102} {\bibfield  {journal} {\bibinfo  {journal}
  {Phys. Rev. Lett.}\ }\textbf {\bibinfo {volume} {117}},\ \bibinfo {pages}
  {101102} (\bibinfo {year} {2016})},\ \Eprint
  {http://arxiv.org/abs/1605.09286} {arXiv:1605.09286 [gr-qc]} \BibitemShut
  {NoStop}%
\bibitem [{\citenamefont {Barausse}\ \emph {et~al.}(2016)\citenamefont
  {Barausse}, \citenamefont {Yunes},\ and\ \citenamefont
  {Chamberlain}}]{Barausse:2016eii}%
  \BibitemOpen
  \bibfield  {author} {\bibinfo {author} {\bibfnamefont {E.}~\bibnamefont
  {Barausse}}, \bibinfo {author} {\bibfnamefont {N.}~\bibnamefont {Yunes}}, \
  and\ \bibinfo {author} {\bibfnamefont {K.}~\bibnamefont {Chamberlain}},\
  }\href {\doibase 10.1103/PhysRevLett.116.241104} {\bibfield  {journal}
  {\bibinfo  {journal} {Phys. Rev. Lett.}\ }\textbf {\bibinfo {volume} {116}},\
  \bibinfo {pages} {241104} (\bibinfo {year} {2016})},\ \Eprint
  {http://arxiv.org/abs/1603.04075} {arXiv:1603.04075 [gr-qc]} \BibitemShut
  {NoStop}%
\bibitem [{\citenamefont {Carson}\ and\ \citenamefont
  {Yagi}(2020)}]{Carson:2019rda}%
  \BibitemOpen
  \bibfield  {author} {\bibinfo {author} {\bibfnamefont {Z.}~\bibnamefont
  {Carson}}\ and\ \bibinfo {author} {\bibfnamefont {K.}~\bibnamefont {Yagi}},\
  }\href {\doibase 10.1088/1361-6382/ab5c9a} {\bibfield  {journal} {\bibinfo
  {journal} {Class. Quant. Grav.}\ }\textbf {\bibinfo {volume} {37}},\ \bibinfo
  {pages} {02LT01} (\bibinfo {year} {2020})},\ \Eprint
  {http://arxiv.org/abs/1905.13155} {arXiv:1905.13155 [gr-qc]} \BibitemShut
  {NoStop}%
\bibitem [{\citenamefont {Gupta}\ \emph {et~al.}(2020)\citenamefont {Gupta},
  \citenamefont {Datta}, \citenamefont {Kastha}, \citenamefont {Borhanian},
  \citenamefont {Arun},\ and\ \citenamefont {Sathyaprakash}}]{Gupta:2020lxa}%
  \BibitemOpen
  \bibfield  {author} {\bibinfo {author} {\bibfnamefont {A.}~\bibnamefont
  {Gupta}}, \bibinfo {author} {\bibfnamefont {S.}~\bibnamefont {Datta}},
  \bibinfo {author} {\bibfnamefont {S.}~\bibnamefont {Kastha}}, \bibinfo
  {author} {\bibfnamefont {S.}~\bibnamefont {Borhanian}}, \bibinfo {author}
  {\bibfnamefont {K.~G.}\ \bibnamefont {Arun}}, \ and\ \bibinfo {author}
  {\bibfnamefont {B.~S.}\ \bibnamefont {Sathyaprakash}},\ }\href {\doibase
  10.1103/PhysRevLett.125.201101} {\bibfield  {journal} {\bibinfo  {journal}
  {Phys. Rev. Lett.}\ }\textbf {\bibinfo {volume} {125}},\ \bibinfo {pages}
  {201101} (\bibinfo {year} {2020})},\ \Eprint
  {http://arxiv.org/abs/2005.09607} {arXiv:2005.09607 [gr-qc]} \BibitemShut
  {NoStop}%
\bibitem [{\citenamefont {Gair}\ \emph {et~al.}(2013)\citenamefont {Gair},
  \citenamefont {Vallisneri}, \citenamefont {Larson},\ and\ \citenamefont
  {Baker}}]{Gair:2012nm}%
  \BibitemOpen
  \bibfield  {author} {\bibinfo {author} {\bibfnamefont {J.~R.}\ \bibnamefont
  {Gair}}, \bibinfo {author} {\bibfnamefont {M.}~\bibnamefont {Vallisneri}},
  \bibinfo {author} {\bibfnamefont {S.~L.}\ \bibnamefont {Larson}}, \ and\
  \bibinfo {author} {\bibfnamefont {J.~G.}\ \bibnamefont {Baker}},\ }\href
  {\doibase 10.12942/lrr-2013-7} {\bibfield  {journal} {\bibinfo  {journal}
  {Living Rev. Rel.}\ }\textbf {\bibinfo {volume} {16}},\ \bibinfo {pages} {7}
  (\bibinfo {year} {2013})},\ \Eprint {http://arxiv.org/abs/1212.5575}
  {arXiv:1212.5575 [gr-qc]} \BibitemShut {NoStop}%
\bibitem [{\citenamefont {Lamberts}\ \emph {et~al.}(2019)\citenamefont
  {Lamberts}, \citenamefont {Blunt}, \citenamefont {Littenberg}, \citenamefont
  {Garrison-Kimmel}, \citenamefont {Kupfer},\ and\ \citenamefont
  {Sanderson}}]{Lamberts_2019}%
  \BibitemOpen
  \bibfield  {author} {\bibinfo {author} {\bibfnamefont {A.}~\bibnamefont
  {Lamberts}}, \bibinfo {author} {\bibfnamefont {S.}~\bibnamefont {Blunt}},
  \bibinfo {author} {\bibfnamefont {T.~B.}\ \bibnamefont {Littenberg}},
  \bibinfo {author} {\bibfnamefont {S.}~\bibnamefont {Garrison-Kimmel}},
  \bibinfo {author} {\bibfnamefont {T.}~\bibnamefont {Kupfer}}, \ and\ \bibinfo
  {author} {\bibfnamefont {R.~E.}\ \bibnamefont {Sanderson}},\ }\href {\doibase
  10.1093/mnras/stz2834} {\bibfield  {journal} {\bibinfo  {journal} {Mon. Not.
  R. Astron. Soc.}\ }\textbf {\bibinfo {volume} {490}},\ \bibinfo {pages}
  {5888} (\bibinfo {year} {2019})}\BibitemShut {NoStop}%
\bibitem [{\citenamefont {{Littenberg}}\ and\ \citenamefont
  {{Yunes}}(2019)}]{Littenberg_2019}%
  \BibitemOpen
  \bibfield  {author} {\bibinfo {author} {\bibfnamefont {T.~B.}\ \bibnamefont
  {{Littenberg}}}\ and\ \bibinfo {author} {\bibfnamefont {N.}~\bibnamefont
  {{Yunes}}},\ }\href {\doibase 10.1088/1361-6382/ab0a3d} {\bibfield  {journal}
  {\bibinfo  {journal} {Classical Quantum Gravity}\ }\textbf {\bibinfo {volume}
  {36}},\ \bibinfo {eid} {095017} (\bibinfo {year} {2019})}\BibitemShut
  {NoStop}%
\bibitem [{\citenamefont {Barbieri}\ \emph {et~al.}(2023)\citenamefont
  {Barbieri}, \citenamefont {Savastano}, \citenamefont {Speri}, \citenamefont
  {Antonelli}, \citenamefont {Sberna}, \citenamefont {Burke}, \citenamefont
  {Gair},\ and\ \citenamefont {Tamanini}}]{Barbieri_2023}%
  \BibitemOpen
  \bibfield  {author} {\bibinfo {author} {\bibfnamefont {R.}~\bibnamefont
  {Barbieri}}, \bibinfo {author} {\bibfnamefont {S.}~\bibnamefont {Savastano}},
  \bibinfo {author} {\bibfnamefont {L.}~\bibnamefont {Speri}}, \bibinfo
  {author} {\bibfnamefont {A.}~\bibnamefont {Antonelli}}, \bibinfo {author}
  {\bibfnamefont {L.}~\bibnamefont {Sberna}}, \bibinfo {author} {\bibfnamefont
  {O.}~\bibnamefont {Burke}}, \bibinfo {author} {\bibfnamefont
  {J.}~\bibnamefont {Gair}}, \ and\ \bibinfo {author} {\bibfnamefont
  {N.}~\bibnamefont {Tamanini}},\ }\href {\doibase 10.1103/PhysRevD.107.064073}
  {\bibfield  {journal} {\bibinfo  {journal} {Phys. Rev. D}\ }\textbf {\bibinfo
  {volume} {107}},\ \bibinfo {pages} {064073} (\bibinfo {year}
  {2023})}\BibitemShut {NoStop}%
\bibitem [{\citenamefont {Kawamura}\ \emph {et~al.}(2008)\citenamefont
  {Kawamura}, \citenamefont {Ando}, \citenamefont {Nakamura}, \citenamefont
  {Tsubono}, \citenamefont {Tanaka}, \citenamefont {Funaki}, \citenamefont
  {Seto}, \citenamefont {Numata}, \citenamefont {Sato}, \citenamefont {Ioka}
  \emph {et~al.}}]{Kawamura_2008}%
  \BibitemOpen
  \bibfield  {author} {\bibinfo {author} {\bibfnamefont {S.}~\bibnamefont
  {Kawamura}}, \bibinfo {author} {\bibfnamefont {M.}~\bibnamefont {Ando}},
  \bibinfo {author} {\bibfnamefont {T.}~\bibnamefont {Nakamura}}, \bibinfo
  {author} {\bibfnamefont {K.}~\bibnamefont {Tsubono}}, \bibinfo {author}
  {\bibfnamefont {T.}~\bibnamefont {Tanaka}}, \bibinfo {author} {\bibfnamefont
  {I.}~\bibnamefont {Funaki}}, \bibinfo {author} {\bibfnamefont
  {N.}~\bibnamefont {Seto}}, \bibinfo {author} {\bibfnamefont {K.}~\bibnamefont
  {Numata}}, \bibinfo {author} {\bibfnamefont {S.}~\bibnamefont {Sato}},
  \bibinfo {author} {\bibfnamefont {K.}~\bibnamefont {Ioka}},  \emph {et~al.},\
  }\href {\doibase 10.1088/1742-6596/122/1/012006} {\bibfield  {journal}
  {\bibinfo  {journal} {J. Phys. Conf. Ser.}\ }\textbf {\bibinfo {volume}
  {122}},\ \bibinfo {pages} {012006} (\bibinfo {year} {2008})}\BibitemShut
  {NoStop}%
\bibitem [{\citenamefont {Poisson}(1998)}]{Poisson_1998}%
  \BibitemOpen
  \bibfield  {author} {\bibinfo {author} {\bibfnamefont {E.}~\bibnamefont
  {Poisson}},\ }\href {\doibase 10.1103/PhysRevD.57.5287} {\bibfield  {journal}
  {\bibinfo  {journal} {Phys. Rev. D}\ }\textbf {\bibinfo {volume} {57}},\
  \bibinfo {pages} {5287} (\bibinfo {year} {1998})}\BibitemShut {NoStop}%
\bibitem [{\citenamefont {Benacquista}(2011)}]{Benacquista_2011}%
  \BibitemOpen
  \bibfield  {author} {\bibinfo {author} {\bibfnamefont {M.~J.}\ \bibnamefont
  {Benacquista}},\ }\href {\doibase 10.1088/2041-8205/740/2/L54} {\bibfield
  {journal} {\bibinfo  {journal} {Astrophys. J.}\ }\textbf {\bibinfo {volume}
  {740}},\ \bibinfo {pages} {L54} (\bibinfo {year} {2011})}\BibitemShut
  {NoStop}%
\bibitem [{\citenamefont {Lai}(2012)}]{Lai_2012}%
  \BibitemOpen
  \bibfield  {author} {\bibinfo {author} {\bibfnamefont {D.}~\bibnamefont
  {Lai}},\ }\href {\doibase 10.1088/2041-8205/757/1/L3} {\bibfield  {journal}
  {\bibinfo  {journal} {Astrophys. J. Lett.}\ }\textbf {\bibinfo {volume}
  {757}},\ \bibinfo {pages} {L3} (\bibinfo {year} {2012})}\BibitemShut
  {NoStop}%
\bibitem [{\citenamefont {Yunes}\ and\ \citenamefont
  {Pretorius}(2009)}]{Yunes_2009}%
  \BibitemOpen
  \bibfield  {author} {\bibinfo {author} {\bibfnamefont {N.}~\bibnamefont
  {Yunes}}\ and\ \bibinfo {author} {\bibfnamefont {F.}~\bibnamefont
  {Pretorius}},\ }\href {\doibase 10.1103/PhysRevD.80.122003} {\bibfield
  {journal} {\bibinfo  {journal} {Phys. Rev. D}\ }\textbf {\bibinfo {volume}
  {80}},\ \bibinfo {pages} {122003} (\bibinfo {year} {2009})}\BibitemShut
  {NoStop}%
\bibitem [{\citenamefont {Yunes}\ and\ \citenamefont
  {Hughes}(2010)}]{Yunes_2010}%
  \BibitemOpen
  \bibfield  {author} {\bibinfo {author} {\bibfnamefont {N.}~\bibnamefont
  {Yunes}}\ and\ \bibinfo {author} {\bibfnamefont {S.~A.}\ \bibnamefont
  {Hughes}},\ }\href {\doibase 10.1103/PhysRevD.82.082002} {\bibfield
  {journal} {\bibinfo  {journal} {Phys. Rev. D}\ }\textbf {\bibinfo {volume}
  {82}},\ \bibinfo {pages} {082002} (\bibinfo {year} {2010})}\BibitemShut
  {NoStop}%
\bibitem [{\citenamefont {Tahura}\ and\ \citenamefont
  {Yagi}(2018)}]{Tahura:2018zuq}%
  \BibitemOpen
  \bibfield  {author} {\bibinfo {author} {\bibfnamefont {S.}~\bibnamefont
  {Tahura}}\ and\ \bibinfo {author} {\bibfnamefont {K.}~\bibnamefont {Yagi}},\
  }\href {\doibase 10.1103/PhysRevD.98.084042} {\bibfield  {journal} {\bibinfo
  {journal} {Phys. Rev. D}\ }\textbf {\bibinfo {volume} {98}},\ \bibinfo
  {pages} {084042} (\bibinfo {year} {2018})},\ \bibinfo {note} {[Erratum:
  Phys.Rev.D 101, 109902 (2020)]},\ \Eprint {http://arxiv.org/abs/1809.00259}
  {arXiv:1809.00259 [gr-qc]} \BibitemShut {NoStop}%
\bibitem [{\citenamefont {Marsh}\ \emph {et~al.}(2004)\citenamefont {Marsh},
  \citenamefont {Nelemans},\ and\ \citenamefont {Steeghs}}]{marsh_2004}%
  \BibitemOpen
  \bibfield  {author} {\bibinfo {author} {\bibfnamefont {T.~R.}\ \bibnamefont
  {Marsh}}, \bibinfo {author} {\bibfnamefont {G.}~\bibnamefont {Nelemans}}, \
  and\ \bibinfo {author} {\bibfnamefont {D.}~\bibnamefont {Steeghs}},\ }\href
  {\doibase 10.1111/j.1365-2966.2004.07564.x} {\bibfield  {journal} {\bibinfo
  {journal} {Mon. Not. R. Astron. Soc.}\ }\textbf {\bibinfo {volume} {350}},\
  \bibinfo {pages} {113} (\bibinfo {year} {2004})},\ \Eprint
  {http://arxiv.org/abs/https://academic.oup.com/mnras/article-pdf/350/1/113/3089404/350-1-113.pdf}
  {https://academic.oup.com/mnras/article-pdf/350/1/113/3089404/350-1-113.pdf}
  \BibitemShut {NoStop}%
\bibitem [{\citenamefont {Yagi}\ and\ \citenamefont
  {Yunes}(2013{\natexlab{a}})}]{Yagi:2013bca}%
  \BibitemOpen
  \bibfield  {author} {\bibinfo {author} {\bibfnamefont {K.}~\bibnamefont
  {Yagi}}\ and\ \bibinfo {author} {\bibfnamefont {N.}~\bibnamefont {Yunes}},\
  }\href {\doibase 10.1126/science.1236462} {\bibfield  {journal} {\bibinfo
  {journal} {Science}\ }\textbf {\bibinfo {volume} {341}},\ \bibinfo {pages}
  {365} (\bibinfo {year} {2013}{\natexlab{a}})},\ \Eprint
  {http://arxiv.org/abs/1302.4499} {arXiv:1302.4499 [gr-qc]} \BibitemShut
  {NoStop}%
\bibitem [{\citenamefont {Yagi}\ and\ \citenamefont
  {Yunes}(2013{\natexlab{b}})}]{Yagi:2013awa}%
  \BibitemOpen
  \bibfield  {author} {\bibinfo {author} {\bibfnamefont {K.}~\bibnamefont
  {Yagi}}\ and\ \bibinfo {author} {\bibfnamefont {N.}~\bibnamefont {Yunes}},\
  }\href {\doibase 10.1103/PhysRevD.88.023009} {\bibfield  {journal} {\bibinfo
  {journal} {Phys. Rev. D}\ }\textbf {\bibinfo {volume} {88}},\ \bibinfo
  {pages} {023009} (\bibinfo {year} {2013}{\natexlab{b}})},\ \Eprint
  {http://arxiv.org/abs/1303.1528} {arXiv:1303.1528 [gr-qc]} \BibitemShut
  {NoStop}%
\bibitem [{\citenamefont {Yagi}\ and\ \citenamefont
  {Yunes}(2017)}]{Yagi:2016bkt}%
  \BibitemOpen
  \bibfield  {author} {\bibinfo {author} {\bibfnamefont {K.}~\bibnamefont
  {Yagi}}\ and\ \bibinfo {author} {\bibfnamefont {N.}~\bibnamefont {Yunes}},\
  }\href {\doibase 10.1016/j.physrep.2017.03.002} {\bibfield  {journal}
  {\bibinfo  {journal} {Phys. Rept.}\ }\textbf {\bibinfo {volume} {681}},\
  \bibinfo {pages} {1} (\bibinfo {year} {2017})},\ \Eprint
  {http://arxiv.org/abs/1608.02582} {arXiv:1608.02582 [gr-qc]} \BibitemShut
  {NoStop}%
\bibitem [{\citenamefont {Boshkayev}\ \emph {et~al.}(2016)\citenamefont
  {Boshkayev}, \citenamefont {Quevedo},\ and\ \citenamefont
  {Zhami}}]{Boshkayev_2017}%
  \BibitemOpen
  \bibfield  {author} {\bibinfo {author} {\bibfnamefont {K.}~\bibnamefont
  {Boshkayev}}, \bibinfo {author} {\bibfnamefont {H.}~\bibnamefont {Quevedo}},
  \ and\ \bibinfo {author} {\bibfnamefont {B.}~\bibnamefont {Zhami}},\ }\href
  {\doibase 10.1093/mnras/stw2614} {\bibfield  {journal} {\bibinfo  {journal}
  {Mon. Not. R. Astron. Soc.}\ }\textbf {\bibinfo {volume} {464}},\ \bibinfo
  {pages} {4349} (\bibinfo {year} {2016})},\ \Eprint
  {http://arxiv.org/abs/https://academic.oup.com/mnras/article-pdf/464/4/4349/8313645/stw2614.pdf}
  {https://academic.oup.com/mnras/article-pdf/464/4/4349/8313645/stw2614.pdf}
  \BibitemShut {NoStop}%
\bibitem [{\citenamefont {{Boshkayev}}\ \emph {et~al.}(2014)\citenamefont
  {{Boshkayev}}, \citenamefont {{Quevedo}}, \citenamefont {{Kalymova}},\ and\
  \citenamefont {{Zhami}}}]{Boshkayev_2014}%
  \BibitemOpen
  \bibfield  {author} {\bibinfo {author} {\bibfnamefont {K.}~\bibnamefont
  {{Boshkayev}}}, \bibinfo {author} {\bibfnamefont {H.}~\bibnamefont
  {{Quevedo}}}, \bibinfo {author} {\bibfnamefont {Z.}~\bibnamefont
  {{Kalymova}}}, \ and\ \bibinfo {author} {\bibfnamefont {B.}~\bibnamefont
  {{Zhami}}},\ }\href {\doibase 10.48550/arXiv.1409.2472} {\bibfield  {journal}
  {\bibinfo  {journal} {arXiv e-prints}\ ,\ \bibinfo {eid} {arXiv:1409.2472}}
  (\bibinfo {year} {2014})},\ \Eprint {http://arxiv.org/abs/1409.2472}
  {arXiv:1409.2472 [astro-ph.SR]} \BibitemShut {NoStop}%
\bibitem [{\citenamefont {Flanagan}\ and\ \citenamefont
  {Hinderer}(2008)}]{Flanagan_08}%
  \BibitemOpen
  \bibfield  {author} {\bibinfo {author} {\bibfnamefont {E.~E.}\ \bibnamefont
  {Flanagan}}\ and\ \bibinfo {author} {\bibfnamefont {T.}~\bibnamefont
  {Hinderer}},\ }\href {\doibase 10.1103/PhysRevD.77.021502} {\bibfield
  {journal} {\bibinfo  {journal} {Phys. Rev. D}\ }\textbf {\bibinfo {volume}
  {77}},\ \bibinfo {pages} {021502} (\bibinfo {year} {2008})}\BibitemShut
  {NoStop}%
\bibitem [{\citenamefont {Wolz}\ \emph {et~al.}(2020)\citenamefont {Wolz},
  \citenamefont {Yagi}, \citenamefont {Anderson},\ and\ \citenamefont
  {Taylor}}]{wolz_2021}%
  \BibitemOpen
  \bibfield  {author} {\bibinfo {author} {\bibfnamefont {A.}~\bibnamefont
  {Wolz}}, \bibinfo {author} {\bibfnamefont {K.}~\bibnamefont {Yagi}}, \bibinfo
  {author} {\bibfnamefont {N.}~\bibnamefont {Anderson}}, \ and\ \bibinfo
  {author} {\bibfnamefont {A.~J.}\ \bibnamefont {Taylor}},\ }\href {\doibase
  10.1093/mnrasl/slaa183} {\bibfield  {journal} {\bibinfo  {journal} {Mon. Not.
  R. Astron. Soc.: Lett.}\ }\textbf {\bibinfo {volume} {500}},\ \bibinfo
  {pages} {L52} (\bibinfo {year} {2020})},\ \Eprint
  {http://arxiv.org/abs/https://academic.oup.com/mnrasl/article-pdf/500/1/L52/34545852/slaa183.pdf}
  {https://academic.oup.com/mnrasl/article-pdf/500/1/L52/34545852/slaa183.pdf}
  \BibitemShut {NoStop}%
\bibitem [{\citenamefont {Wade}\ \emph {et~al.}(2014)\citenamefont {Wade},
  \citenamefont {Creighton}, \citenamefont {Ochsner}, \citenamefont {Lackey},
  \citenamefont {Farr}, \citenamefont {Littenberg},\ and\ \citenamefont
  {Raymond}}]{Wade_2014}%
  \BibitemOpen
  \bibfield  {author} {\bibinfo {author} {\bibfnamefont {L.}~\bibnamefont
  {Wade}}, \bibinfo {author} {\bibfnamefont {J.~D.~E.}\ \bibnamefont
  {Creighton}}, \bibinfo {author} {\bibfnamefont {E.}~\bibnamefont {Ochsner}},
  \bibinfo {author} {\bibfnamefont {B.~D.}\ \bibnamefont {Lackey}}, \bibinfo
  {author} {\bibfnamefont {B.~F.}\ \bibnamefont {Farr}}, \bibinfo {author}
  {\bibfnamefont {T.~B.}\ \bibnamefont {Littenberg}}, \ and\ \bibinfo {author}
  {\bibfnamefont {V.}~\bibnamefont {Raymond}},\ }\href {\doibase
  10.1103/PhysRevD.89.103012} {\bibfield  {journal} {\bibinfo  {journal} {Phys.
  Rev. D}\ }\textbf {\bibinfo {volume} {89}},\ \bibinfo {pages} {103012}
  (\bibinfo {year} {2014})}\BibitemShut {NoStop}%
\bibitem [{\citenamefont {Ioka}\ and\ \citenamefont
  {Taniguchi}(2000)}]{Ioka_2000}%
  \BibitemOpen
  \bibfield  {author} {\bibinfo {author} {\bibfnamefont {K.}~\bibnamefont
  {Ioka}}\ and\ \bibinfo {author} {\bibfnamefont {K.}~\bibnamefont
  {Taniguchi}},\ }\href {\doibase 10.1086/309004} {\bibfield  {journal}
  {\bibinfo  {journal} {Astrophys. J.}\ }\textbf {\bibinfo {volume} {537}},\
  \bibinfo {pages} {327} (\bibinfo {year} {2000})}\BibitemShut {NoStop}%
\bibitem [{\citenamefont {Keresztes}\ \emph {et~al.}(2005)\citenamefont
  {Keresztes}, \citenamefont {Mik\'oczi},\ and\ \citenamefont
  {Gergely}}]{Keresztes_2005}%
  \BibitemOpen
  \bibfield  {author} {\bibinfo {author} {\bibfnamefont {Z.}~\bibnamefont
  {Keresztes}}, \bibinfo {author} {\bibfnamefont {B.}~\bibnamefont
  {Mik\'oczi}}, \ and\ \bibinfo {author} {\bibfnamefont {L.~A.}\ \bibnamefont
  {Gergely}},\ }\href {\doibase 10.1103/PhysRevD.72.104022} {\bibfield
  {journal} {\bibinfo  {journal} {Phys. Rev. D}\ }\textbf {\bibinfo {volume}
  {72}},\ \bibinfo {pages} {104022} (\bibinfo {year} {2005})}\BibitemShut
  {NoStop}%
\bibitem [{\citenamefont {Henry}\ \emph {et~al.}(2024)\citenamefont {Henry},
  \citenamefont {Larrouturou},\ and\ \citenamefont
  {Le~Poncin-Lafitte}}]{Henry:2023len}%
  \BibitemOpen
  \bibfield  {author} {\bibinfo {author} {\bibfnamefont {Q.}~\bibnamefont
  {Henry}}, \bibinfo {author} {\bibfnamefont {F.}~\bibnamefont {Larrouturou}},
  \ and\ \bibinfo {author} {\bibfnamefont {C.}~\bibnamefont
  {Le~Poncin-Lafitte}},\ }\href {\doibase 10.1103/PhysRevD.109.084048}
  {\bibfield  {journal} {\bibinfo  {journal} {Phys. Rev. D}\ }\textbf {\bibinfo
  {volume} {109}},\ \bibinfo {pages} {084048} (\bibinfo {year} {2024})},\
  \Eprint {http://arxiv.org/abs/2310.03785} {arXiv:2310.03785 [gr-qc]}
  \BibitemShut {NoStop}%
\bibitem [{\citenamefont {{Goldreich}}\ and\ \citenamefont
  {{Lynden-Bell}}(1969)}]{Goldreich_1969}%
  \BibitemOpen
  \bibfield  {author} {\bibinfo {author} {\bibfnamefont {P.}~\bibnamefont
  {{Goldreich}}}\ and\ \bibinfo {author} {\bibfnamefont {D.}~\bibnamefont
  {{Lynden-Bell}}},\ }\href {\doibase 10.1086/149947} {\bibfield  {journal}
  {\bibinfo  {journal} {Astrophys.~J.}\ }\textbf {\bibinfo {volume} {156}},\
  \bibinfo {pages} {59} (\bibinfo {year} {1969})}\BibitemShut {NoStop}%
\bibitem [{\citenamefont {Lidov}(1962)}]{Lidov_1962}%
  \BibitemOpen
  \bibfield  {author} {\bibinfo {author} {\bibfnamefont {M.}~\bibnamefont
  {Lidov}},\ }\href {\doibase https://doi.org/10.1016/0032-0633(62)90129-0}
  {\bibfield  {journal} {\bibinfo  {journal} {Planet. Space Sci.}\ }\textbf
  {\bibinfo {volume} {9}},\ \bibinfo {pages} {719} (\bibinfo {year}
  {1962})}\BibitemShut {NoStop}%
\bibitem [{\citenamefont {{Kozai}}(1962)}]{Kozai_1962}%
  \BibitemOpen
  \bibfield  {author} {\bibinfo {author} {\bibfnamefont {Y.}~\bibnamefont
  {{Kozai}}},\ }\href {\doibase 10.1086/108790} {\bibfield  {journal} {\bibinfo
   {journal} {Astrophys.~J.}\ }\textbf {\bibinfo {volume} {67}},\ \bibinfo
  {pages} {591} (\bibinfo {year} {1962})}\BibitemShut {NoStop}%
\bibitem [{\citenamefont {{Verbunt}}\ and\ \citenamefont
  {{Rappaport}}(1988)}]{Verbunt_1988}%
  \BibitemOpen
  \bibfield  {author} {\bibinfo {author} {\bibfnamefont {F.}~\bibnamefont
  {{Verbunt}}}\ and\ \bibinfo {author} {\bibfnamefont {S.}~\bibnamefont
  {{Rappaport}}},\ }\href {\doibase 10.1086/166645} {\bibfield  {journal}
  {\bibinfo  {journal} {Astrophys.~J.~}\ }\textbf {\bibinfo {volume} {332}},\
  \bibinfo {pages} {193} (\bibinfo {year} {1988})}\BibitemShut {NoStop}%
\bibitem [{\citenamefont {{Biscoveanu}}\ \emph {et~al.}(2023)\citenamefont
  {{Biscoveanu}}, \citenamefont {{Kremer}},\ and\ \citenamefont
  {{Thrane}}}]{Biscoveanu_2023}%
  \BibitemOpen
  \bibfield  {author} {\bibinfo {author} {\bibfnamefont {S.}~\bibnamefont
  {{Biscoveanu}}}, \bibinfo {author} {\bibfnamefont {K.}~\bibnamefont
  {{Kremer}}}, \ and\ \bibinfo {author} {\bibfnamefont {E.}~\bibnamefont
  {{Thrane}}},\ }\href {\doibase 10.3847/1538-4357/acc585} {\bibfield
  {journal} {\bibinfo  {journal} {Astrophys. J.}\ }\textbf {\bibinfo {volume}
  {949}},\ \bibinfo {eid} {95} (\bibinfo {year} {2023})},\ \Eprint
  {http://arxiv.org/abs/2206.15390} {arXiv:2206.15390 [astro-ph.HE]}
  \BibitemShut {NoStop}%
\bibitem [{\citenamefont {Yi}\ \emph {et~al.}(2024)\citenamefont {Yi},
  \citenamefont {Lau}, \citenamefont {Yagi},\ and\ \citenamefont
  {Arras}}]{Yi:2023osk}%
  \BibitemOpen
  \bibfield  {author} {\bibinfo {author} {\bibfnamefont {S.}~\bibnamefont
  {Yi}}, \bibinfo {author} {\bibfnamefont {S.~Y.}\ \bibnamefont {Lau}},
  \bibinfo {author} {\bibfnamefont {K.}~\bibnamefont {Yagi}}, \ and\ \bibinfo
  {author} {\bibfnamefont {P.}~\bibnamefont {Arras}},\ }\href {\doibase
  10.1093/mnras/stae1453} {\bibfield  {journal} {\bibinfo  {journal} {Mon. Not.
  Roy. Astron. Soc.}\ }\textbf {\bibinfo {volume} {531}},\ \bibinfo {pages}
  {4681} (\bibinfo {year} {2024})},\ \Eprint {http://arxiv.org/abs/2310.16172}
  {arXiv:2310.16172 [astro-ph.HE]} \BibitemShut {NoStop}%
\bibitem [{\citenamefont {Kidder}\ \emph {et~al.}(1993)\citenamefont {Kidder},
  \citenamefont {Will},\ and\ \citenamefont {Wiseman}}]{Kidder_1993}%
  \BibitemOpen
  \bibfield  {author} {\bibinfo {author} {\bibfnamefont {L.~E.}\ \bibnamefont
  {Kidder}}, \bibinfo {author} {\bibfnamefont {C.~M.}\ \bibnamefont {Will}}, \
  and\ \bibinfo {author} {\bibfnamefont {A.~G.}\ \bibnamefont {Wiseman}},\
  }\href {\doibase 10.1103/PhysRevD.47.R4183} {\bibfield  {journal} {\bibinfo
  {journal} {Phys. Rev. D}\ }\textbf {\bibinfo {volume} {47}},\ \bibinfo
  {pages} {R4183} (\bibinfo {year} {1993})}\BibitemShut {NoStop}%
\bibitem [{\citenamefont {{Burdge}}\ \emph {et~al.}(2019)\citenamefont
  {{Burdge}}, \citenamefont {{Coughlin}}, \citenamefont {{Fuller}},
  \citenamefont {{Kupfer}}, \citenamefont {{Bellm}}, \citenamefont
  {{Bildsten}}, \citenamefont {{Graham}}, \citenamefont {{Kaplan}},
  \citenamefont {{Roestel}}, \citenamefont {{Dekany}}, \citenamefont {{Duev}},
  \citenamefont {{Feeney}}, \citenamefont {{Giomi}}, \citenamefont {{Helou}},
  \citenamefont {{Kaye}}, \citenamefont {{Laher}}, \citenamefont {{Mahabal}},
  \citenamefont {{Masci}}, \citenamefont {{Riddle}}, \citenamefont {{Shupe}},
  \citenamefont {{Soumagnac}}, \citenamefont {{Smith}}, \citenamefont
  {{Szkody}}, \citenamefont {{Walters}}, \citenamefont {{Kulkarni}},\ and\
  \citenamefont {{Prince}}}]{Burdge_2019}%
  \BibitemOpen
  \bibfield  {author} {\bibinfo {author} {\bibfnamefont {K.~B.}\ \bibnamefont
  {{Burdge}}}, \bibinfo {author} {\bibfnamefont {M.~W.}\ \bibnamefont
  {{Coughlin}}}, \bibinfo {author} {\bibfnamefont {J.}~\bibnamefont
  {{Fuller}}}, \bibinfo {author} {\bibfnamefont {T.}~\bibnamefont {{Kupfer}}},
  \bibinfo {author} {\bibfnamefont {E.~C.}\ \bibnamefont {{Bellm}}}, \bibinfo
  {author} {\bibfnamefont {L.}~\bibnamefont {{Bildsten}}}, \bibinfo {author}
  {\bibfnamefont {M.~J.}\ \bibnamefont {{Graham}}}, \bibinfo {author}
  {\bibfnamefont {D.~L.}\ \bibnamefont {{Kaplan}}}, \bibinfo {author}
  {\bibfnamefont {J.~v.}\ \bibnamefont {{Roestel}}}, \bibinfo {author}
  {\bibfnamefont {R.~G.}\ \bibnamefont {{Dekany}}}, \bibinfo {author}
  {\bibfnamefont {D.~A.}\ \bibnamefont {{Duev}}}, \bibinfo {author}
  {\bibfnamefont {M.}~\bibnamefont {{Feeney}}}, \bibinfo {author}
  {\bibfnamefont {M.}~\bibnamefont {{Giomi}}}, \bibinfo {author} {\bibfnamefont
  {G.}~\bibnamefont {{Helou}}}, \bibinfo {author} {\bibfnamefont
  {S.}~\bibnamefont {{Kaye}}}, \bibinfo {author} {\bibfnamefont {R.~R.}\
  \bibnamefont {{Laher}}}, \bibinfo {author} {\bibfnamefont {A.~A.}\
  \bibnamefont {{Mahabal}}}, \bibinfo {author} {\bibfnamefont {F.~J.}\
  \bibnamefont {{Masci}}}, \bibinfo {author} {\bibfnamefont {R.}~\bibnamefont
  {{Riddle}}}, \bibinfo {author} {\bibfnamefont {D.~L.}\ \bibnamefont
  {{Shupe}}}, \bibinfo {author} {\bibfnamefont {M.~T.}\ \bibnamefont
  {{Soumagnac}}}, \bibinfo {author} {\bibfnamefont {R.~M.}\ \bibnamefont
  {{Smith}}}, \bibinfo {author} {\bibfnamefont {P.}~\bibnamefont {{Szkody}}},
  \bibinfo {author} {\bibfnamefont {R.}~\bibnamefont {{Walters}}}, \bibinfo
  {author} {\bibfnamefont {S.~R.}\ \bibnamefont {{Kulkarni}}}, \ and\ \bibinfo
  {author} {\bibfnamefont {T.~A.}\ \bibnamefont {{Prince}}},\ }\href {\doibase
  10.1038/s41586-019-1403-0} {\bibfield  {journal} {\bibinfo  {journal}
  {Nature}\ }\textbf {\bibinfo {volume} {571}},\ \bibinfo {pages} {528}
  (\bibinfo {year} {2019})},\ \Eprint {http://arxiv.org/abs/1907.11291}
  {arXiv:1907.11291 [astro-ph.SR]} \BibitemShut {NoStop}%
\bibitem [{\citenamefont {Cornish}\ and\ \citenamefont
  {Rubbo}(2003)}]{Cornish_2003}%
  \BibitemOpen
  \bibfield  {author} {\bibinfo {author} {\bibfnamefont {N.~J.}\ \bibnamefont
  {Cornish}}\ and\ \bibinfo {author} {\bibfnamefont {L.~J.}\ \bibnamefont
  {Rubbo}},\ }\href {\doibase 10.1103/PhysRevD.67.022001} {\bibfield  {journal}
  {\bibinfo  {journal} {Phys. Rev. D}\ }\textbf {\bibinfo {volume} {67}},\
  \bibinfo {pages} {022001} (\bibinfo {year} {2003})}\BibitemShut {NoStop}%
\bibitem [{\citenamefont {Cutler}(1998)}]{PhysRevD.57.7089}%
  \BibitemOpen
  \bibfield  {author} {\bibinfo {author} {\bibfnamefont {C.}~\bibnamefont
  {Cutler}},\ }\href {\doibase 10.1103/PhysRevD.57.7089} {\bibfield  {journal}
  {\bibinfo  {journal} {Phys. Rev. D}\ }\textbf {\bibinfo {volume} {57}},\
  \bibinfo {pages} {7089} (\bibinfo {year} {1998})}\BibitemShut {NoStop}%
\bibitem [{\citenamefont {Barack}\ and\ \citenamefont
  {Cutler}(2004)}]{PhysRevD.69.082005}%
  \BibitemOpen
  \bibfield  {author} {\bibinfo {author} {\bibfnamefont {L.}~\bibnamefont
  {Barack}}\ and\ \bibinfo {author} {\bibfnamefont {C.}~\bibnamefont
  {Cutler}},\ }\href {\doibase 10.1103/PhysRevD.69.082005} {\bibfield
  {journal} {\bibinfo  {journal} {Phys. Rev. D}\ }\textbf {\bibinfo {volume}
  {69}},\ \bibinfo {pages} {082005} (\bibinfo {year} {2004})}\BibitemShut
  {NoStop}%
\bibitem [{\citenamefont {{Shah, S.}}\ \emph {et~al.}(2012)\citenamefont
  {{Shah, S.}}, \citenamefont {{van der Sluys, M.}},\ and\ \citenamefont
  {{Nelemans, G.}}}]{Shah_2012}%
  \BibitemOpen
  \bibfield  {author} {\bibinfo {author} {\bibnamefont {{Shah, S.}}}, \bibinfo
  {author} {\bibnamefont {{van der Sluys, M.}}}, \ and\ \bibinfo {author}
  {\bibnamefont {{Nelemans, G.}}},\ }\href {\doibase
  10.1051/0004-6361/201219309} {\bibfield  {journal} {\bibinfo  {journal}
  {A\&A}\ }\textbf {\bibinfo {volume} {544}},\ \bibinfo {pages} {A153}
  (\bibinfo {year} {2012})}\BibitemShut {NoStop}%
\bibitem [{\citenamefont {Takahashi}\ and\ \citenamefont
  {Seto}(2002)}]{Takahashi_2002}%
  \BibitemOpen
  \bibfield  {author} {\bibinfo {author} {\bibfnamefont {R.}~\bibnamefont
  {Takahashi}}\ and\ \bibinfo {author} {\bibfnamefont {N.}~\bibnamefont
  {Seto}},\ }\href {\doibase 10.1086/341483} {\bibfield  {journal} {\bibinfo
  {journal} {Astrophys.~J.}\ }\textbf {\bibinfo {volume} {575}},\ \bibinfo
  {pages} {1030} (\bibinfo {year} {2002})}\BibitemShut {NoStop}%
\bibitem [{\citenamefont {Robson}\ \emph {et~al.}(2019)\citenamefont {Robson},
  \citenamefont {Cornish},\ and\ \citenamefont {Liu}}]{Robson_2019}%
  \BibitemOpen
  \bibfield  {author} {\bibinfo {author} {\bibfnamefont {T.}~\bibnamefont
  {Robson}}, \bibinfo {author} {\bibfnamefont {N.~J.}\ \bibnamefont {Cornish}},
  \ and\ \bibinfo {author} {\bibfnamefont {C.}~\bibnamefont {Liu}},\ }\href
  {\doibase 10.1088/1361-6382/ab1101} {\bibfield  {journal} {\bibinfo
  {journal} {Classical Quantum Gravity}\ }\textbf {\bibinfo {volume} {36}},\
  \bibinfo {pages} {105011} (\bibinfo {year} {2019})}\BibitemShut {NoStop}%
\bibitem [{\citenamefont {Yunes}\ \emph
  {et~al.}(2016{\natexlab{b}})\citenamefont {Yunes}, \citenamefont {Yagi},\
  and\ \citenamefont {Pretorius}}]{Yunes_2016}%
  \BibitemOpen
  \bibfield  {author} {\bibinfo {author} {\bibfnamefont {N.}~\bibnamefont
  {Yunes}}, \bibinfo {author} {\bibfnamefont {K.}~\bibnamefont {Yagi}}, \ and\
  \bibinfo {author} {\bibfnamefont {F.}~\bibnamefont {Pretorius}},\ }\href
  {\doibase 10.1103/PhysRevD.94.084002} {\bibfield  {journal} {\bibinfo
  {journal} {Phys. Rev. D}\ }\textbf {\bibinfo {volume} {94}},\ \bibinfo
  {pages} {084002} (\bibinfo {year} {2016}{\natexlab{b}})}\BibitemShut
  {NoStop}%
\bibitem [{\citenamefont {Burdge}\ \emph {et~al.}(2020)\citenamefont {Burdge},
  \citenamefont {Prince}, \citenamefont {Fuller}, \citenamefont {Kaplan},
  \citenamefont {Marsh}, \citenamefont {Tremblay}, \citenamefont {Zhuang},
  \citenamefont {Bellm}, \citenamefont {Caiazzo}, \citenamefont {Coughlin}
  \emph {et~al.}}]{Burdge_2020}%
  \BibitemOpen
  \bibfield  {author} {\bibinfo {author} {\bibfnamefont {K.~B.}\ \bibnamefont
  {Burdge}}, \bibinfo {author} {\bibfnamefont {T.~A.}\ \bibnamefont {Prince}},
  \bibinfo {author} {\bibfnamefont {J.}~\bibnamefont {Fuller}}, \bibinfo
  {author} {\bibfnamefont {D.~L.}\ \bibnamefont {Kaplan}}, \bibinfo {author}
  {\bibfnamefont {T.~R.}\ \bibnamefont {Marsh}}, \bibinfo {author}
  {\bibfnamefont {P.-E.}\ \bibnamefont {Tremblay}}, \bibinfo {author}
  {\bibfnamefont {Z.}~\bibnamefont {Zhuang}}, \bibinfo {author} {\bibfnamefont
  {E.~C.}\ \bibnamefont {Bellm}}, \bibinfo {author} {\bibfnamefont
  {I.}~\bibnamefont {Caiazzo}}, \bibinfo {author} {\bibfnamefont {M.~W.}\
  \bibnamefont {Coughlin}},  \emph {et~al.},\ }\href {\doibase
  10.3847/1538-4357/abc261} {\bibfield  {journal} {\bibinfo  {journal}
  {Astrophys.~J.}\ }\textbf {\bibinfo {volume} {905}},\ \bibinfo {pages} {32}
  (\bibinfo {year} {2020})}\BibitemShut {NoStop}%
\bibitem [{\citenamefont {Hermes}\ \emph {et~al.}(2012)\citenamefont {Hermes},
  \citenamefont {Kilic}, \citenamefont {Brown}, \citenamefont {Winget},
  \citenamefont {Prieto}, \citenamefont {Gianninas}, \citenamefont {Mukadam},
  \citenamefont {Cabrera-Lavers},\ and\ \citenamefont {Kenyon}}]{Hermes_2012}%
  \BibitemOpen
  \bibfield  {author} {\bibinfo {author} {\bibfnamefont {J.~J.}\ \bibnamefont
  {Hermes}}, \bibinfo {author} {\bibfnamefont {M.}~\bibnamefont {Kilic}},
  \bibinfo {author} {\bibfnamefont {W.~R.}\ \bibnamefont {Brown}}, \bibinfo
  {author} {\bibfnamefont {D.~E.}\ \bibnamefont {Winget}}, \bibinfo {author}
  {\bibfnamefont {C.~A.}\ \bibnamefont {Prieto}}, \bibinfo {author}
  {\bibfnamefont {A.}~\bibnamefont {Gianninas}}, \bibinfo {author}
  {\bibfnamefont {A.~S.}\ \bibnamefont {Mukadam}}, \bibinfo {author}
  {\bibfnamefont {A.}~\bibnamefont {Cabrera-Lavers}}, \ and\ \bibinfo {author}
  {\bibfnamefont {S.~J.}\ \bibnamefont {Kenyon}},\ }\href {\doibase
  10.1088/2041-8205/757/2/L21} {\bibfield  {journal} {\bibinfo  {journal}
  {Astrophys. J. Lett.}\ }\textbf {\bibinfo {volume} {757}},\ \bibinfo {pages}
  {L21} (\bibinfo {year} {2012})}\BibitemShut {NoStop}%
\bibitem [{\citenamefont {Kinugawa}\ \emph {et~al.}(2022)\citenamefont
  {Kinugawa}, \citenamefont {Takeda}, \citenamefont {Tanikawa},\ and\
  \citenamefont {Yamaguchi}}]{Kinugawa:2019uey}%
  \BibitemOpen
  \bibfield  {author} {\bibinfo {author} {\bibfnamefont {T.}~\bibnamefont
  {Kinugawa}}, \bibinfo {author} {\bibfnamefont {H.}~\bibnamefont {Takeda}},
  \bibinfo {author} {\bibfnamefont {A.}~\bibnamefont {Tanikawa}}, \ and\
  \bibinfo {author} {\bibfnamefont {H.}~\bibnamefont {Yamaguchi}},\ }\href
  {\doibase 10.3847/1538-4357/ac9135} {\bibfield  {journal} {\bibinfo
  {journal} {Astrophys. J.}\ }\textbf {\bibinfo {volume} {938}},\ \bibinfo
  {pages} {52} (\bibinfo {year} {2022})},\ \Eprint
  {http://arxiv.org/abs/1910.01063} {arXiv:1910.01063 [astro-ph.HE]}
  \BibitemShut {NoStop}%
\bibitem [{\citenamefont {Poisson}\ and\ \citenamefont
  {Will}(1995)}]{Poisson:1995ef}%
  \BibitemOpen
  \bibfield  {author} {\bibinfo {author} {\bibfnamefont {E.}~\bibnamefont
  {Poisson}}\ and\ \bibinfo {author} {\bibfnamefont {C.~M.}\ \bibnamefont
  {Will}},\ }\href {\doibase 10.1103/PhysRevD.52.848} {\bibfield  {journal}
  {\bibinfo  {journal} {Phys. Rev. D}\ }\textbf {\bibinfo {volume} {52}},\
  \bibinfo {pages} {848} (\bibinfo {year} {1995})},\ \Eprint
  {http://arxiv.org/abs/gr-qc/9502040} {arXiv:gr-qc/9502040} \BibitemShut
  {NoStop}%
\bibitem [{\citenamefont {Cutler}\ and\ \citenamefont
  {Vallisneri}(2007)}]{Cutler_2007}%
  \BibitemOpen
  \bibfield  {author} {\bibinfo {author} {\bibfnamefont {C.}~\bibnamefont
  {Cutler}}\ and\ \bibinfo {author} {\bibfnamefont {M.}~\bibnamefont
  {Vallisneri}},\ }\href {\doibase 10.1103/PhysRevD.76.104018} {\bibfield
  {journal} {\bibinfo  {journal} {Phys. Rev. D}\ }\textbf {\bibinfo {volume}
  {76}},\ \bibinfo {pages} {104018} (\bibinfo {year} {2007})}\BibitemShut
  {NoStop}%
\bibitem [{\citenamefont {Zhang}\ \emph {et~al.}(2017)\citenamefont {Zhang},
  \citenamefont {Liu},\ and\ \citenamefont {Zhao}}]{Zhang_2017}%
  \BibitemOpen
  \bibfield  {author} {\bibinfo {author} {\bibfnamefont {X.}~\bibnamefont
  {Zhang}}, \bibinfo {author} {\bibfnamefont {T.}~\bibnamefont {Liu}}, \ and\
  \bibinfo {author} {\bibfnamefont {W.}~\bibnamefont {Zhao}},\ }\href {\doibase
  10.1103/PhysRevD.95.104027} {\bibfield  {journal} {\bibinfo  {journal} {Phys.
  Rev. D}\ }\textbf {\bibinfo {volume} {95}},\ \bibinfo {pages} {104027}
  (\bibinfo {year} {2017})}\BibitemShut {NoStop}%
\bibitem [{\citenamefont {Zhang}\ \emph {et~al.}(2019)\citenamefont {Zhang},
  \citenamefont {Zhao}, \citenamefont {Liu}, \citenamefont {Lin}, \citenamefont
  {Zhang}, \citenamefont {Zhao}, \citenamefont {Zhang}, \citenamefont {Zhu},\
  and\ \citenamefont {Wang}}]{Zhang_2019}%
  \BibitemOpen
  \bibfield  {author} {\bibinfo {author} {\bibfnamefont {X.}~\bibnamefont
  {Zhang}}, \bibinfo {author} {\bibfnamefont {W.}~\bibnamefont {Zhao}},
  \bibinfo {author} {\bibfnamefont {T.}~\bibnamefont {Liu}}, \bibinfo {author}
  {\bibfnamefont {K.}~\bibnamefont {Lin}}, \bibinfo {author} {\bibfnamefont
  {C.}~\bibnamefont {Zhang}}, \bibinfo {author} {\bibfnamefont
  {X.}~\bibnamefont {Zhao}}, \bibinfo {author} {\bibfnamefont {S.}~\bibnamefont
  {Zhang}}, \bibinfo {author} {\bibfnamefont {T.}~\bibnamefont {Zhu}}, \ and\
  \bibinfo {author} {\bibfnamefont {A.}~\bibnamefont {Wang}},\ }\href {\doibase
  10.3847/1538-4357/ab09f4} {\bibfield  {journal} {\bibinfo  {journal}
  {Astrophys. J.}\ }\textbf {\bibinfo {volume} {874}},\ \bibinfo {pages} {121}
  (\bibinfo {year} {2019})}\BibitemShut {NoStop}%
\bibitem [{\citenamefont {Freire}\ \emph {et~al.}(2012)\citenamefont {Freire},
  \citenamefont {Wex}, \citenamefont {Esposito-Farèse}, \citenamefont
  {Verbiest}, \citenamefont {Bailes}, \citenamefont {Jacoby}, \citenamefont
  {Kramer}, \citenamefont {Stairs}, \citenamefont {Antoniadis},\ and\
  \citenamefont {Janssen}}]{Freire_2012}%
  \BibitemOpen
  \bibfield  {author} {\bibinfo {author} {\bibfnamefont {P.~C.~C.}\
  \bibnamefont {Freire}}, \bibinfo {author} {\bibfnamefont {N.}~\bibnamefont
  {Wex}}, \bibinfo {author} {\bibfnamefont {G.}~\bibnamefont
  {Esposito-Farèse}}, \bibinfo {author} {\bibfnamefont {J.~P.~W.}\
  \bibnamefont {Verbiest}}, \bibinfo {author} {\bibfnamefont {M.}~\bibnamefont
  {Bailes}}, \bibinfo {author} {\bibfnamefont {B.~A.}\ \bibnamefont {Jacoby}},
  \bibinfo {author} {\bibfnamefont {M.}~\bibnamefont {Kramer}}, \bibinfo
  {author} {\bibfnamefont {I.~H.}\ \bibnamefont {Stairs}}, \bibinfo {author}
  {\bibfnamefont {J.}~\bibnamefont {Antoniadis}}, \ and\ \bibinfo {author}
  {\bibfnamefont {G.~H.}\ \bibnamefont {Janssen}},\ }\href {\doibase
  10.1111/j.1365-2966.2012.21253.x} {\bibfield  {journal} {\bibinfo  {journal}
  {Mon. Not. R. Astron. Soc.}\ }\textbf {\bibinfo {volume} {423}},\ \bibinfo
  {pages} {3328} (\bibinfo {year} {2012})},\ \Eprint
  {http://arxiv.org/abs/https://academic.oup.com/mnras/article-pdf/423/4/3328/18196694/mnras0423-3328.pdf}
  {https://academic.oup.com/mnras/article-pdf/423/4/3328/18196694/mnras0423-3328.pdf}
  \BibitemShut {NoStop}%
\bibitem [{\citenamefont {Zhang}\ \emph {et~al.}()\citenamefont {Zhang},
  \citenamefont {Zhao}, \citenamefont {Liu}, \citenamefont {Lin}, \citenamefont
  {Zhang}, \citenamefont {Zhang}, \citenamefont {Zhao}, \citenamefont {Zhu},\
  and\ \citenamefont {Wang}}]{Zhang_2018}%
  \BibitemOpen
  \bibfield  {author} {\bibinfo {author} {\bibfnamefont {X.}~\bibnamefont
  {Zhang}}, \bibinfo {author} {\bibfnamefont {W.}~\bibnamefont {Zhao}},
  \bibinfo {author} {\bibfnamefont {T.}~\bibnamefont {Liu}}, \bibinfo {author}
  {\bibfnamefont {K.}~\bibnamefont {Lin}}, \bibinfo {author} {\bibfnamefont
  {C.}~\bibnamefont {Zhang}}, \bibinfo {author} {\bibfnamefont
  {S.}~\bibnamefont {Zhang}}, \bibinfo {author} {\bibfnamefont
  {X.}~\bibnamefont {Zhao}}, \bibinfo {author} {\bibfnamefont {T.}~\bibnamefont
  {Zhu}}, \ and\ \bibinfo {author} {\bibfnamefont {A.}~\bibnamefont {Wang}},\
  }\href {https://arxiv.org/abs/1806.02581} {\ }\Eprint
  {http://arxiv.org/abs/1806.02581} {arXiv:1806.02581 [gr-qc]} \BibitemShut
  {NoStop}%
\bibitem [{\citenamefont {Laarakkers}\ and\ \citenamefont
  {Poisson}(1999)}]{Laarakkers_1999}%
  \BibitemOpen
  \bibfield  {author} {\bibinfo {author} {\bibfnamefont {W.~G.}\ \bibnamefont
  {Laarakkers}}\ and\ \bibinfo {author} {\bibfnamefont {E.}~\bibnamefont
  {Poisson}},\ }\href {\doibase 10.1086/306732} {\bibfield  {journal} {\bibinfo
   {journal} {Astrophys.~J.}\ }\textbf {\bibinfo {volume} {512}},\ \bibinfo
  {pages} {282–287} (\bibinfo {year} {1999})}\BibitemShut {NoStop}%
\bibitem [{\citenamefont {Amaro-Seoane}(2019)}]{Amaro-Seoane_2019}%
  \BibitemOpen
  \bibfield  {author} {\bibinfo {author} {\bibfnamefont {P.}~\bibnamefont
  {Amaro-Seoane}},\ }\href {\doibase 10.1103/PhysRevD.99.123025} {\bibfield
  {journal} {\bibinfo  {journal} {Phys. Rev. D}\ }\textbf {\bibinfo {volume}
  {99}},\ \bibinfo {pages} {123025} (\bibinfo {year} {2019})}\BibitemShut
  {NoStop}%
\bibitem [{\citenamefont {Hook}\ and\ \citenamefont
  {Huang}(2018)}]{Hook:2017psm}%
  \BibitemOpen
  \bibfield  {author} {\bibinfo {author} {\bibfnamefont {A.}~\bibnamefont
  {Hook}}\ and\ \bibinfo {author} {\bibfnamefont {J.}~\bibnamefont {Huang}},\
  }\href {\doibase 10.1007/JHEP06(2018)036} {\bibfield  {journal} {\bibinfo
  {journal} {JHEP}\ }\textbf {\bibinfo {volume} {06}},\ \bibinfo {pages} {036}
  (\bibinfo {year} {2018})},\ \Eprint {http://arxiv.org/abs/1708.08464}
  {arXiv:1708.08464 [hep-ph]} \BibitemShut {NoStop}%
\bibitem [{\citenamefont {Huang}\ \emph
  {et~al.}(2019{\natexlab{a}})\citenamefont {Huang}, \citenamefont {Johnson},
  \citenamefont {Sagunski}, \citenamefont {Sakellariadou},\ and\ \citenamefont
  {Zhang}}]{Huang:2018pbu}%
  \BibitemOpen
  \bibfield  {author} {\bibinfo {author} {\bibfnamefont {J.}~\bibnamefont
  {Huang}}, \bibinfo {author} {\bibfnamefont {M.~C.}\ \bibnamefont {Johnson}},
  \bibinfo {author} {\bibfnamefont {L.}~\bibnamefont {Sagunski}}, \bibinfo
  {author} {\bibfnamefont {M.}~\bibnamefont {Sakellariadou}}, \ and\ \bibinfo
  {author} {\bibfnamefont {J.}~\bibnamefont {Zhang}},\ }\href {\doibase
  10.1103/PhysRevD.99.063013} {\bibfield  {journal} {\bibinfo  {journal} {Phys.
  Rev. D}\ }\textbf {\bibinfo {volume} {99}},\ \bibinfo {pages} {063013}
  (\bibinfo {year} {2019}{\natexlab{a}})},\ \Eprint
  {http://arxiv.org/abs/1807.02133} {arXiv:1807.02133 [hep-ph]} \BibitemShut
  {NoStop}%
\bibitem [{\citenamefont {Huang}\ \emph
  {et~al.}(2019{\natexlab{b}})\citenamefont {Huang}, \citenamefont {Johnson},
  \citenamefont {Sagunski}, \citenamefont {Sakellariadou},\ and\ \citenamefont
  {Zhang}}]{Huang_2019}%
  \BibitemOpen
  \bibfield  {author} {\bibinfo {author} {\bibfnamefont {J.}~\bibnamefont
  {Huang}}, \bibinfo {author} {\bibfnamefont {M.~C.}\ \bibnamefont {Johnson}},
  \bibinfo {author} {\bibfnamefont {L.}~\bibnamefont {Sagunski}}, \bibinfo
  {author} {\bibfnamefont {M.}~\bibnamefont {Sakellariadou}}, \ and\ \bibinfo
  {author} {\bibfnamefont {J.}~\bibnamefont {Zhang}},\ }\href {\doibase
  10.1103/PhysRevD.99.063013} {\bibfield  {journal} {\bibinfo  {journal} {Phys.
  Rev. D}\ }\textbf {\bibinfo {volume} {99}},\ \bibinfo {pages} {063013}
  (\bibinfo {year} {2019}{\natexlab{b}})}\BibitemShut {NoStop}%
\bibitem [{\citenamefont {Peccei}\ and\ \citenamefont
  {Quinn}(1977)}]{Peccei_1977}%
  \BibitemOpen
  \bibfield  {author} {\bibinfo {author} {\bibfnamefont {R.~D.}\ \bibnamefont
  {Peccei}}\ and\ \bibinfo {author} {\bibfnamefont {H.~R.}\ \bibnamefont
  {Quinn}},\ }\href {\doibase 10.1103/PhysRevLett.38.1440} {\bibfield
  {journal} {\bibinfo  {journal} {Phys. Rev. Lett.}\ }\textbf {\bibinfo
  {volume} {38}},\ \bibinfo {pages} {1440} (\bibinfo {year}
  {1977})}\BibitemShut {NoStop}%
\bibitem [{\citenamefont {Weinberg}(1978)}]{Weinberg_1978}%
  \BibitemOpen
  \bibfield  {author} {\bibinfo {author} {\bibfnamefont {S.}~\bibnamefont
  {Weinberg}},\ }\href {\doibase 10.1103/PhysRevLett.40.223} {\bibfield
  {journal} {\bibinfo  {journal} {Phys. Rev. Lett.}\ }\textbf {\bibinfo
  {volume} {40}},\ \bibinfo {pages} {223} (\bibinfo {year} {1978})}\BibitemShut
  {NoStop}%
\bibitem [{\citenamefont {Wilczek}(1978)}]{Wilczek_1978}%
  \BibitemOpen
  \bibfield  {author} {\bibinfo {author} {\bibfnamefont {F.}~\bibnamefont
  {Wilczek}},\ }\href {\doibase 10.1103/PhysRevLett.40.279} {\bibfield
  {journal} {\bibinfo  {journal} {Phys. Rev. Lett.}\ }\textbf {\bibinfo
  {volume} {40}},\ \bibinfo {pages} {279} (\bibinfo {year} {1978})}\BibitemShut
  {NoStop}%
\bibitem [{\citenamefont {Marsh}(2016)}]{Marsh_2016}%
  \BibitemOpen
  \bibfield  {author} {\bibinfo {author} {\bibfnamefont {D.~J.}\ \bibnamefont
  {Marsh}},\ }\href {\doibase https://doi.org/10.1016/j.physrep.2016.06.005}
  {\bibfield  {journal} {\bibinfo  {journal} {Phys. Rep.}\ }\textbf {\bibinfo
  {volume} {643}},\ \bibinfo {pages} {1} (\bibinfo {year} {2016})}\BibitemShut
  {NoStop}%
\bibitem [{\citenamefont {Preskill}\ \emph {et~al.}(1983)\citenamefont
  {Preskill}, \citenamefont {Wise},\ and\ \citenamefont
  {Wilczek}}]{Preskill_1983}%
  \BibitemOpen
  \bibfield  {author} {\bibinfo {author} {\bibfnamefont {J.}~\bibnamefont
  {Preskill}}, \bibinfo {author} {\bibfnamefont {M.~B.}\ \bibnamefont {Wise}},
  \ and\ \bibinfo {author} {\bibfnamefont {F.}~\bibnamefont {Wilczek}},\ }\href
  {\doibase https://doi.org/10.1016/0370-2693(83)90637-8} {\bibfield  {journal}
  {\bibinfo  {journal} {Phys. Lett. B}\ }\textbf {\bibinfo {volume} {120}},\
  \bibinfo {pages} {127} (\bibinfo {year} {1983})}\BibitemShut {NoStop}%
\bibitem [{\citenamefont {Arik}\ \emph {et~al.}(2011)\citenamefont {Arik} \emph
  {et~al.}}]{Arik_2011}%
  \BibitemOpen
  \bibfield  {author} {\bibinfo {author} {\bibfnamefont {M.}~\bibnamefont
  {Arik}} \emph {et~al.} (\bibinfo {collaboration} {CAST Collaboration}),\
  }\href {\doibase 10.1103/PhysRevLett.107.261302} {\bibfield  {journal}
  {\bibinfo  {journal} {Phys. Rev. Lett.}\ }\textbf {\bibinfo {volume} {107}},\
  \bibinfo {pages} {261302} (\bibinfo {year} {2011})}\BibitemShut {NoStop}%
\bibitem [{\citenamefont {Asztalos}\ \emph {et~al.}(2004)\citenamefont
  {Asztalos}, \citenamefont {Bradley}, \citenamefont {Duffy}, \citenamefont
  {Hagmann}, \citenamefont {Kinion}, \citenamefont {Moltz}, \citenamefont
  {Rosenberg}, \citenamefont {Sikivie}, \citenamefont {Stoeffl}, \citenamefont
  {Sullivan}, \citenamefont {Tanner}, \citenamefont {van Bibber},\ and\
  \citenamefont {Yu}}]{Asztalos_2004}%
  \BibitemOpen
  \bibfield  {author} {\bibinfo {author} {\bibfnamefont {S.~J.}\ \bibnamefont
  {Asztalos}}, \bibinfo {author} {\bibfnamefont {R.~F.}\ \bibnamefont
  {Bradley}}, \bibinfo {author} {\bibfnamefont {L.}~\bibnamefont {Duffy}},
  \bibinfo {author} {\bibfnamefont {C.}~\bibnamefont {Hagmann}}, \bibinfo
  {author} {\bibfnamefont {D.}~\bibnamefont {Kinion}}, \bibinfo {author}
  {\bibfnamefont {D.~M.}\ \bibnamefont {Moltz}}, \bibinfo {author}
  {\bibfnamefont {L.~J.}\ \bibnamefont {Rosenberg}}, \bibinfo {author}
  {\bibfnamefont {P.}~\bibnamefont {Sikivie}}, \bibinfo {author} {\bibfnamefont
  {W.}~\bibnamefont {Stoeffl}}, \bibinfo {author} {\bibfnamefont {N.~S.}\
  \bibnamefont {Sullivan}}, \bibinfo {author} {\bibfnamefont {D.~B.}\
  \bibnamefont {Tanner}}, \bibinfo {author} {\bibfnamefont {K.}~\bibnamefont
  {van Bibber}}, \ and\ \bibinfo {author} {\bibfnamefont {D.~B.}\ \bibnamefont
  {Yu}},\ }\href {\doibase 10.1103/PhysRevD.69.011101} {\bibfield  {journal}
  {\bibinfo  {journal} {Phys. Rev. D}\ }\textbf {\bibinfo {volume} {69}},\
  \bibinfo {pages} {011101} (\bibinfo {year} {2004})}\BibitemShut {NoStop}%
\bibitem [{\citenamefont {Budker}\ \emph {et~al.}(2014)\citenamefont {Budker},
  \citenamefont {Graham}, \citenamefont {Ledbetter}, \citenamefont
  {Rajendran},\ and\ \citenamefont {Sushkov}}]{Budker_2014}%
  \BibitemOpen
  \bibfield  {author} {\bibinfo {author} {\bibfnamefont {D.}~\bibnamefont
  {Budker}}, \bibinfo {author} {\bibfnamefont {P.~W.}\ \bibnamefont {Graham}},
  \bibinfo {author} {\bibfnamefont {M.}~\bibnamefont {Ledbetter}}, \bibinfo
  {author} {\bibfnamefont {S.}~\bibnamefont {Rajendran}}, \ and\ \bibinfo
  {author} {\bibfnamefont {A.~O.}\ \bibnamefont {Sushkov}},\ }\href {\doibase
  10.1103/PhysRevX.4.021030} {\bibfield  {journal} {\bibinfo  {journal} {Phys.
  Rev. X}\ }\textbf {\bibinfo {volume} {4}},\ \bibinfo {pages} {021030}
  (\bibinfo {year} {2014})}\BibitemShut {NoStop}%
\bibitem [{\citenamefont {Raffelt}(1986)}]{Raffelt_1986}%
  \BibitemOpen
  \bibfield  {author} {\bibinfo {author} {\bibfnamefont {G.~G.}\ \bibnamefont
  {Raffelt}},\ }\href {\doibase 10.1103/PhysRevD.33.897} {\bibfield  {journal}
  {\bibinfo  {journal} {Phys. Rev. D}\ }\textbf {\bibinfo {volume} {33}},\
  \bibinfo {pages} {897} (\bibinfo {year} {1986})}\BibitemShut {NoStop}%
\bibitem [{\citenamefont {Ellis}\ and\ \citenamefont
  {Olive}(1987)}]{Ellis_1987}%
  \BibitemOpen
  \bibfield  {author} {\bibinfo {author} {\bibfnamefont {J.}~\bibnamefont
  {Ellis}}\ and\ \bibinfo {author} {\bibfnamefont {K.}~\bibnamefont {Olive}},\
  }\href {\doibase https://doi.org/10.1016/0370-2693(87)91710-2} {\bibfield
  {journal} {\bibinfo  {journal} {Phys. Lett. B}\ }\textbf {\bibinfo {volume}
  {193}},\ \bibinfo {pages} {525} (\bibinfo {year} {1987})}\BibitemShut
  {NoStop}%
\bibitem [{\citenamefont {Janka}\ \emph {et~al.}(1996)\citenamefont {Janka},
  \citenamefont {Keil}, \citenamefont {Raffelt},\ and\ \citenamefont
  {Seckel}}]{Janka_1996}%
  \BibitemOpen
  \bibfield  {author} {\bibinfo {author} {\bibfnamefont {H.-T.}\ \bibnamefont
  {Janka}}, \bibinfo {author} {\bibfnamefont {W.}~\bibnamefont {Keil}},
  \bibinfo {author} {\bibfnamefont {G.}~\bibnamefont {Raffelt}}, \ and\
  \bibinfo {author} {\bibfnamefont {D.}~\bibnamefont {Seckel}},\ }\href
  {\doibase 10.1103/PhysRevLett.76.2621} {\bibfield  {journal} {\bibinfo
  {journal} {Phys. Rev. Lett.}\ }\textbf {\bibinfo {volume} {76}},\ \bibinfo
  {pages} {2621} (\bibinfo {year} {1996})}\BibitemShut {NoStop}%
\bibitem [{\citenamefont {Poddar}\ \emph {et~al.}(2020)\citenamefont {Poddar},
  \citenamefont {Mohanty},\ and\ \citenamefont {Jana}}]{Poddar_2020}%
  \BibitemOpen
  \bibfield  {author} {\bibinfo {author} {\bibfnamefont {T.~K.}\ \bibnamefont
  {Poddar}}, \bibinfo {author} {\bibfnamefont {S.}~\bibnamefont {Mohanty}}, \
  and\ \bibinfo {author} {\bibfnamefont {S.}~\bibnamefont {Jana}},\ }\href
  {\doibase 10.1103/PhysRevD.101.083007} {\bibfield  {journal} {\bibinfo
  {journal} {Phys. Rev. D}\ }\textbf {\bibinfo {volume} {101}},\ \bibinfo
  {pages} {083007} (\bibinfo {year} {2020})}\BibitemShut {NoStop}%
\bibitem [{\citenamefont {Antoniadis}\ \emph {et~al.}(2013)\citenamefont
  {Antoniadis}, \citenamefont {Freire}, \citenamefont {Wex}, \citenamefont
  {Tauris}, \citenamefont {Lynch}, \citenamefont {van Kerkwijk}, \citenamefont
  {Kramer}, \citenamefont {Bassa}, \citenamefont {Dhillon}, \citenamefont
  {Driebe}, \citenamefont {Hessels}, \citenamefont {Kaspi}, \citenamefont
  {Kondratiev}, \citenamefont {Langer}, \citenamefont {Marsh}, \citenamefont
  {McLaughlin}, \citenamefont {Pennucci}, \citenamefont {Ransom}, \citenamefont
  {Stairs}, \citenamefont {van Leeuwen}, \citenamefont {Verbiest},\ and\
  \citenamefont {Whelan}}]{Antoniadis_2013}%
  \BibitemOpen
  \bibfield  {author} {\bibinfo {author} {\bibfnamefont {J.}~\bibnamefont
  {Antoniadis}}, \bibinfo {author} {\bibfnamefont {P.~C.~C.}\ \bibnamefont
  {Freire}}, \bibinfo {author} {\bibfnamefont {N.}~\bibnamefont {Wex}},
  \bibinfo {author} {\bibfnamefont {T.~M.}\ \bibnamefont {Tauris}}, \bibinfo
  {author} {\bibfnamefont {R.~S.}\ \bibnamefont {Lynch}}, \bibinfo {author}
  {\bibfnamefont {M.~H.}\ \bibnamefont {van Kerkwijk}}, \bibinfo {author}
  {\bibfnamefont {M.}~\bibnamefont {Kramer}}, \bibinfo {author} {\bibfnamefont
  {C.}~\bibnamefont {Bassa}}, \bibinfo {author} {\bibfnamefont {V.~S.}\
  \bibnamefont {Dhillon}}, \bibinfo {author} {\bibfnamefont {T.}~\bibnamefont
  {Driebe}}, \bibinfo {author} {\bibfnamefont {J.~W.~T.}\ \bibnamefont
  {Hessels}}, \bibinfo {author} {\bibfnamefont {V.~M.}\ \bibnamefont {Kaspi}},
  \bibinfo {author} {\bibfnamefont {V.~I.}\ \bibnamefont {Kondratiev}},
  \bibinfo {author} {\bibfnamefont {N.}~\bibnamefont {Langer}}, \bibinfo
  {author} {\bibfnamefont {T.~R.}\ \bibnamefont {Marsh}}, \bibinfo {author}
  {\bibfnamefont {M.~A.}\ \bibnamefont {McLaughlin}}, \bibinfo {author}
  {\bibfnamefont {T.~T.}\ \bibnamefont {Pennucci}}, \bibinfo {author}
  {\bibfnamefont {S.~M.}\ \bibnamefont {Ransom}}, \bibinfo {author}
  {\bibfnamefont {I.~H.}\ \bibnamefont {Stairs}}, \bibinfo {author}
  {\bibfnamefont {J.}~\bibnamefont {van Leeuwen}}, \bibinfo {author}
  {\bibfnamefont {J.~P.~W.}\ \bibnamefont {Verbiest}}, \ and\ \bibinfo {author}
  {\bibfnamefont {D.~G.}\ \bibnamefont {Whelan}},\ }\href {\doibase
  10.1126/science.1233232} {\bibfield  {journal} {\bibinfo  {journal}
  {Science}\ }\textbf {\bibinfo {volume} {340}},\ \bibinfo {pages} {1233232}
  (\bibinfo {year} {2013})}\BibitemShut {NoStop}%
\bibitem [{\citenamefont {Gasperini}(1987)}]{Gasperini_1987}%
  \BibitemOpen
  \bibfield  {author} {\bibinfo {author} {\bibfnamefont {M.}~\bibnamefont
  {Gasperini}},\ }\href {\doibase 10.1088/0264-9381/4/2/026} {\bibfield
  {journal} {\bibinfo  {journal} {Classical and Quantum Gravity}\ }\textbf
  {\bibinfo {volume} {4}},\ \bibinfo {pages} {485} (\bibinfo {year}
  {1987})}\BibitemShut {NoStop}%
\bibitem [{\citenamefont {Fradkin}\ and\ \citenamefont
  {Tseytlin}(1985)}]{Fradkin_1985}%
  \BibitemOpen
  \bibfield  {author} {\bibinfo {author} {\bibfnamefont {E.}~\bibnamefont
  {Fradkin}}\ and\ \bibinfo {author} {\bibfnamefont {A.}~\bibnamefont
  {Tseytlin}},\ }\href {\doibase https://doi.org/10.1016/0550-3213(85)90559-0}
  {\bibfield  {journal} {\bibinfo  {journal} {Nucl. Phys. B}\ }\textbf
  {\bibinfo {volume} {261}},\ \bibinfo {pages} {1} (\bibinfo {year}
  {1985})}\BibitemShut {NoStop}%
\bibitem [{\citenamefont {Garay}\ and\ \citenamefont
  {García-Bellido}(1993)}]{Garay_1993}%
  \BibitemOpen
  \bibfield  {author} {\bibinfo {author} {\bibfnamefont {L.~J.}\ \bibnamefont
  {Garay}}\ and\ \bibinfo {author} {\bibfnamefont {J.}~\bibnamefont
  {García-Bellido}},\ }\href {\doibase
  https://doi.org/10.1016/0550-3213(93)90411-H} {\bibfield  {journal} {\bibinfo
   {journal} {Nucl. Phys. B}\ }\textbf {\bibinfo {volume} {400}},\ \bibinfo
  {pages} {416} (\bibinfo {year} {1993})}\BibitemShut {NoStop}%
\bibitem [{\citenamefont {Wagoner}(1970)}]{Wagoner_1970}%
  \BibitemOpen
  \bibfield  {author} {\bibinfo {author} {\bibfnamefont {R.~V.}\ \bibnamefont
  {Wagoner}},\ }\href {\doibase 10.1103/PhysRevD.1.3209} {\bibfield  {journal}
  {\bibinfo  {journal} {Phys. Rev. D}\ }\textbf {\bibinfo {volume} {1}},\
  \bibinfo {pages} {3209} (\bibinfo {year} {1970})}\BibitemShut {NoStop}%
\bibitem [{\citenamefont {Isham}\ \emph {et~al.}(1971)\citenamefont {Isham},
  \citenamefont {Salam},\ and\ \citenamefont {Strathdee}}]{Isham_1971}%
  \BibitemOpen
  \bibfield  {author} {\bibinfo {author} {\bibfnamefont {C.~J.}\ \bibnamefont
  {Isham}}, \bibinfo {author} {\bibfnamefont {A.}~\bibnamefont {Salam}}, \ and\
  \bibinfo {author} {\bibfnamefont {J.}~\bibnamefont {Strathdee}},\ }\href
  {\doibase 10.1103/PhysRevD.3.867} {\bibfield  {journal} {\bibinfo  {journal}
  {Phys. Rev. D}\ }\textbf {\bibinfo {volume} {3}},\ \bibinfo {pages} {867}
  (\bibinfo {year} {1971})}\BibitemShut {NoStop}%
\bibitem [{\citenamefont {Hassan}\ and\ \citenamefont
  {Rosen}(2012)}]{Hassan_2011}%
  \BibitemOpen
  \bibfield  {author} {\bibinfo {author} {\bibfnamefont {S.~F.}\ \bibnamefont
  {Hassan}}\ and\ \bibinfo {author} {\bibfnamefont {R.~A.}\ \bibnamefont
  {Rosen}},\ }\href {\doibase 10.1007/JHEP02(2012)126} {\bibfield  {journal}
  {\bibinfo  {journal} {JHEP}\ }\textbf {\bibinfo {volume} {02}},\ \bibinfo
  {pages} {126} (\bibinfo {year} {2012})},\ \Eprint
  {http://arxiv.org/abs/1109.3515} {arXiv:1109.3515 [hep-th]} \BibitemShut
  {NoStop}%
\bibitem [{\citenamefont {Yunes}\ and\ \citenamefont
  {Stein}(2011)}]{Yunes_2011}%
  \BibitemOpen
  \bibfield  {author} {\bibinfo {author} {\bibfnamefont {N.}~\bibnamefont
  {Yunes}}\ and\ \bibinfo {author} {\bibfnamefont {L.~C.}\ \bibnamefont
  {Stein}},\ }\href {\doibase 10.1103/PhysRevD.83.104002} {\bibfield  {journal}
  {\bibinfo  {journal} {Phys. Rev. D}\ }\textbf {\bibinfo {volume} {83}},\
  \bibinfo {pages} {104002} (\bibinfo {year} {2011})}\BibitemShut {NoStop}%
\end{thebibliography}%

\end{document}